\documentstyle[11pt,amssymbols,pb-diagram]{article}
\newcommand{\Matrix}[5]{%
\begin{array}{cc}
#1&#2\\
#3&#4
\end{array}\Big|#5}

\newcommand{\mmatrix}[4]{%
\left(#1\begin{array}{c}#2\\#3\end{array}\Big|#4\right)}

\newcommand{\CL}{{\cal L}}

\newcommand{\mcb}[1]{(#1;x^4,x^{2r})_\infty}
\newcommand{\newwet}[5]{W'\left(\matrix{#1&#2\cr#3&#4\cr}\Big|#5\right)}
\def\ds{~\hbox{${\scriptstyle>\atop\raise3pt\hbox{$\scriptstyle<$}}$}~}

\def\gl{\begin{array}{l}\\[-8mm]{\scriptstyle >}\\[-4mm]{\scriptstyle <} \end{array}}
\def\Ker{{\rm Ker}\,}
\def\Im{{\rm Im}\,}
\def\Res{{\rm Res}\,}
\def\cL{{\cal{L}}}

\def\be{\begin{equation}}
\def\en{\end{equation}}
\def\bea{\begin{eqnarray}}
\def\ena{\end{eqnarray}}
\def\bean{\begin{eqnarray*}}
\def\enan{\end{eqnarray*}}

\def\mref#1{(\ref{#1})}

\def\Z{{\bf Z}}

\def\th{\theta}

\def\z{\zeta}
\def\x{\xi}

\def\bpl#1{{[}{#1}]_x^+}
\def\bra#1{\langle #1 |}	
\def\ket#1{| #1\rangle}		
\def\br#1{\langle #1 \rangle}	

\def\path{|p\rangle}

\def\exp{{\rm exp}\,}

\def\com[#1,#2]{\hbox{$[#1,#2]$}}

\def\pr#1{\left(#1\right)_\infty}  
\def\Phit{\widetilde{\Phi}}  
\def\Psit{\widetilde{\Psi}}  

\def\Phim#1{\mathrel{\mathop{\kern0pt \Phi}\limits^#1}}
\def\Psim#1{\mathrel{\mathop{\kern0pt \Psi}\limits^#1}}
\def\Phin#1{\mathrel{\mathop{\kern0pt \Phit}\limits^#1}}
\def\Psin#1{\mathrel{\mathop{\kern0pt \Psit}\limits^#1}}

\def\be{\begin{equation}}
\def\ee{\end{equation}}
\newcommand{\bqa}{\begin{eqnarray}}  
\newcommand{\eqa}{\end{eqnarray}}  
\newcommand{\ra}{\rightarrow}

\def\G{\Gamma}
\def\half{{1 \over 2}}

\def\a{\alpha}  
\def\b{\beta}  
\def\g{\gamma}  
\def\d{\delta}  
\def\dz{{dz\over 2\pi i z}}

\def\ep{\epsilon} 
 
\def\l{\lambda}

\def\t{\theta}
  
\def\ov{\over}  
\def\pr{\prime}  
  
\def\ed{\end{document}}
\def\ws{\;\;}

\def\ra{\rightarrow}  
  
\def\2pi{1\over 2\pi i}  
  
\def\qu{{1\ov 4}}
\def\at{{\alpha \ov 2}}
\def\newline{\hfil\break}

\def\ra{\rightarrow}

\def\sq2{\sqrt{2}}  
\def\sqk2{\sqrt{2(k+2}}  
\def\sqk{\sqrt{k}}

\def\bfig{\begin{figure}}
\def\bfigt{\begin{figure}[top]}
\def\efig{\end{figure}}
\def\bea{\begin{eqnarray}}  
\def\eea{\end{eqnarray}}  
\def\br{\begin{array}}
\def\er{\end{array}}
\def\ea{\end{array}\end{equation}}
\def\bac{\begin{equation}\begin{array}{rll}}
\def\ba{\begin{equation}\begin{array}}

\def\sg#1#2#3#4#5{\tilde{S}\left(\left.\begin{array}{ll}{#1}&{#2}\\{#3}&{#4}\end{array}\right|{#5}\right)}
\def\bsg#1#2#3{\tilde{M}_{#1}^{({#2},k)}({#3})}
\def\votg{\tilde{\Psi}^{*}}

\def\Mg{\tilde{M}}
\def\we#1#2#3#4#5{W\left(\left.\begin{array}{ll}{#1}&{#2}\\{#3}&{#4}\end{array}\right|{#5}\right)}

\def\wet#1#2#3#4#5{W'\left(\left.\begin{array}{ll}{#1}&{#2}\\{#3}&{#4}\end{array}\right|{#5}\right)}

\def\qp#1#2{({#1}\, ; \, {#2})_{\infty}}
\def\nb#1#2#3#4{K^{({#1})}_{#2}\left(\left.k{#3}{k\atop k} \right| {#4}\right)}
\def\nbp#1#2#3#4{K^{({#1})}_{#2}\left(\left.{#3}{k\atop k} \right|
{#4}\right)}
\def\nbpm#1#2#3#4{K^{({#1})}_{#2}\left(\left.{#3}{k+1\atop k+1} \right|
{#4}\right)}
\def\nbt#1#2#3#4{K'^{\;({#1})}_{#2}\left(\left.{#3}{k\atop k} \right|
{#4}\right)}
\def\bnb#1#2#3#4{\bar{K}^{({#1})}_{#2}\left(\left.k{#3}{k\atop k} \right| {#4}\right)}

\def\lab{\label}
\def\lbs#1{ {_{B}^{c}}\langle {#1},k |}
\def\rbs#1{|{#1},k\rangle_B^c}
\def\lbsb#1{ {_{B}^{c}}\langle \overline{{#1},k} |}
\def\rbsb#1{|\overline{{#1},k}\rangle_B^c}
\def\cbt#1{( {#1};\,x^4,\,x^{2(r-1)} )_{\infty}}
\def\cb#1{({#1};\,x^4,\,x^{2r})_{\infty}}
\def\br#1{({#1};\,x^{2r})_{\infty}}
\def\brt#1{({#1};\,x^{2(r-1)})_{\infty}}

\def\Z{{\Bbb Z}}

\def\cF{{\cal{F}}}
\def\cF{{\cal{F}}}

\def\vo{\Phi}  
\def\vot{\Psi^{*}}

\def\cA{{\cal{A}}}  
\def\cB{{\cal{B}}}

\def\pl{\prod\limits} 
 
\def\plp{\prod\limits_{m>0}} 
\def\sl{\sum\limits} 

\def\slp{\sl_{m>0}}

\def\ep{\varepsilon} 
\def\epp{\varepsilon^{\prime}} 
\def\eppp{\varepsilon^{\prime \prime}}

\def\ss{\subsection}

\def\xib{\bar{\xi}}

\def\z{\zeta}
\def\zi{\zeta^{-1}}
\def\xii{\xi^{-1}}

\begin{document}
\bibliographystyle{unsrt}

\begin{flushright}

RIMS-1106, DTP-96-41\\
\end{flushright}
\begin{center}
{\LARGE Boundary ABF Models }\\[8mm]
{Tetsuji Miwa$\!^1$ and Robert Weston$\!^2$\\[8mm]
October 1996}
\end{center}
\footnotetext[1]{Research Institute for Mathematical Sciences,
Kyoto University, Kyoto 606, Japan.}
\footnotetext[2]{Department of Mathematical Sciences,
University of Durham, Science Labs,
South Rd, Durham DH1 3LE, U.K.\\ \hspace*{5mm} R.A.Weston@durham.ac.uk}

\begin{abstract}
\noindent We diagonalise the transfer matrix of boundary ABF models using
bosonized vertex operators. We compute the boundary S-matrix and 
check the scaling limit against known results for perturbed boundary
conformal field theories.
\end{abstract}
\section{Introduction}
The theory of solvable lattice models, which originated in the work
of Bethe and Onsager in the 30-40's, has matured in the last 20 years.
Various models were found to be solvable and various methods were
invented to solve these models.
In the course of this work, some remarkable interactions occured
between the theory of solvable
lattice models and other branches of mathematics and physics, e.g.,
representation theory and conformal field theory.

The ABF models we are going to study in this paper were introduced in \cite{ABF}.
Baxter's book \cite{Bax82} presented the corner transfer matrix method,
by which the one point correlation functions were computed.
The general results obtained in \cite{ABF} strongly suggested a link between
solvable lattice models and the representation theory of the Virasoro algebra.

Such a link was manifest from the very beginning in conformal field theory.
The representation theory of the Virasoro algebra was very successfully
applied to conformal field theory to obtain correlation
functions and the differential equations which they obeyed.

Solvability of lattice models was understood by means of the method of
commuting transfer matrices. The algebraic structure behind it was
given shape by the idea of $q$-deformation. Finally, in \cite{Daval92,JM94}, 
the corner transfer matrix method was given its correct place in the
representation theory. The link was established between the highest 
weight representations
of the quantum affine algebra $U_q\bigl(\widehat{sl}_2\bigr)$
and the XXZ model in the anti-ferromagnetic regime.
Namely, the spaces of the eigenvectors of the corner
transfer matrix of the latter were identified with the level $1$
highest weight representations of the former, and the half transfer matrices
were identified with the intertwiners of these representations.
In fact, a similar structure had been known in conformal field
theory. The operators of this kind were called vertex operators.
These operators were realized using representation theory \cite{TK}.
The method of bosonization was also very powerful \cite{FF}.

As for the ABF models, such a scheme was only recently used successfully
\cite{LP96}. The bosonization method of conformal
field theory, by which the minimal unitary models were solved,
was deformed to give a realization of the half transfer matrices of the
ABF models.

The boundary problem has been studied in both lattice and continuum theories.
In \cite{JKKKM}, the vertex operator method was extended to the boundary
XXZ model. The aim of this paper is to extend the bosonization method
to the boundary ABF models.

We start from a known solution of the boundary Yang-Baxter equation 
\cite{BPO95}. We then introduce the boundary transfer matrix of Sklyanin-type 
in terms of a product of vertex operators.
Our goal is to diagonalise the boundary transfer matrix.
We shall give an explicit formula for the boundary vacuum state and compute
the boundary $S$-matrix. The bosonization scheme of \cite{LP96}
is based on a cohomological construction. Thus, in the boundary ABF models,
the physical spaces of states are realized as sub-quotients of the bosonic
Fock spaces on which the screening operators act. One of the main points
in our work is to show that the boundary vacuum states belong to
the sub-quotients.

The scaling limit of the bulk ABF models in regime III is 
described by the  $\phi_{(1,3)}$ deformations of the $c=1-6/r(r-1)$ rational
conformal field theories. In  \cite{GZ} and \cite{Ch95}, boundary
S-matrices are computed for the perturbed $r=4$ and $r=5$ models on 
a half line (the Ising and Tri-critical Ising models respectively). 
We check that our boundary S-matrices agree with these results in the
scaling limit.

The plan of the paper is as follows. Section 2 prepares the boundary
Boltzmann weights and the transfer matrix. Section 3 is a summary of the
bosonization scheme. We give the boundary vacuum state in Section 4,
and compute the boundary $S$-matrices in Section 5. Section 6 takes the
scaling limit. Appendix A gives useful formulas for operator products.
Appendix B gives a proof of the eigenvector relation for the
boundary vacuum states.
\setcounter{section}{1}
\setcounter{subsection}{1}
\section{Bulk and boundary weights}
We recall the Boltzmann weights of the ABF model, and
set up the boundary transfer matrix in terms of the
vertex operators.
\subsection{The bulk weights}
The ABF model has two parameters $x$ and $r$.
We assume that $0<x<1$ and $r\geqslant4$
$(r \in {\Bbb Z})$.
We use the symbol $[u]$ for the elliptic theta function.
\begin{eqnarray*}
&& [u] = x^{\frac{u^2}r-u}\Theta_{x^{2r}}(x^{2u}),\\
&& \Theta_q(z) = (z;q)_{\infty}(qz^{-1};q)_{\infty}(q;q)_{\infty},\\
&& (z;q_1,\cdots,q_m)_{\infty}
= \prod_{i_1,\cdots,i_m=0}^{\infty}(1-q_1^{i_1}\cdots q_m^{i_m}z).
\end{eqnarray*}
There are six kinds of configurations around a face and
the corresponding Boltzmann weights are given as follows \cite{ABF}.
\bea
&&W\left(
\begin{array}{cc}
k \pm 1 & k \\
k       & k \mp1 
\end{array}
\Big| u \right)
=R(u),\label{w1}\\
&&W\left(
\begin{array}{cc}
k       & k \pm 1 \\
k \pm 1 & k
\end{array}
\Big| u \right)
=R(u)\frac{[k\pm u][1]}{[1-u][k]},\label{w2}\\
&&W\left(
\begin{array}{cc}
k       & k \pm 1 \\
k \mp 1 & k
\end{array}
\Big| u \right)
=R(u)\frac{[k\mp 1][u]}{[k][1-u]}.\label{w3}
\eea
Here $k$ is an integer such that $1\leqslant k \leqslant r-1$,
and the normalisation factor $R(u)$ is given by
\begin{eqnarray*}
R(u) 
&=& \zeta^{\frac{r-1}{2r}} \frac{\rho(\zeta)}{\rho(\zeta^{-1})},
\quad \zeta=x^{2u},\\
\rho(\zeta)
&=& \frac{
(x^4\zeta; x^4,x^{2r})_{\infty}
(x^{2r}\zeta; x^4,x^{2r})_{\infty}}
{(x^2\zeta; x^4,x^{2r})_{\infty}
(x^{2+2r}\zeta; x^4,x^{2r})_{\infty}}.
\end{eqnarray*}
Graphically, we represent 
$W\left( \Matrix dcabu \right) $ by

\begin{center}
\begin{picture}(0,0)%
\includegraphics{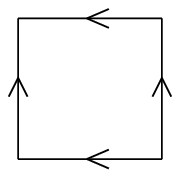}%
\end{picture}%
\setlength{\unitlength}{0.01250000in}%
\begingroup\makeatletter\ifx\SetFigFont\undefined
\def\x#1#2#3#4#5#6#7\relax{\def\x{#1#2#3#4#5#6}}%
\expandafter\x\fmtname xxxxxx\relax \def\y{splain}%
\ifx\x\y   
\gdef\SetFigFont#1#2#3{%
  \ifnum #1<17\tiny\else \ifnum #1<20\small\else
  \ifnum #1<24\normalsize\else \ifnum #1<29\large\else
  \ifnum #1<34\Large\else \ifnum #1<41\LARGE\else
     \huge\fi\fi\fi\fi\fi\fi
  \csname #3\endcsname}%
\else
\gdef\SetFigFont#1#2#3{\begingroup
  \count@#1\relax \ifnum 25<\count@\count@25\fi
  \def\x{\endgroup\@setsize\SetFigFont{#2pt}}%
  \expandafter\x
    \csname \romannumeral\the\count@ pt\expandafter\endcsname
    \csname @\romannumeral\the\count@ pt\endcsname
  \csname #3\endcsname}%
\fi
\fi\endgroup
\begin{picture}(69,76)(311,593)
\put(345,627){\makebox(0,0)[lb]{\smash{\SetFigFont{12}{14.4}{rm}$u$}}}
\put(380,595){\makebox(0,0)[lb]{\smash{\SetFigFont{12}{14.4}{rm}$b$}}}
\put(380,660){\makebox(0,0)[lb]{\smash{\SetFigFont{12}{14.4}{rm}$c$}}}
\put(311,595){\makebox(0,0)[lb]{\smash{\SetFigFont{12}{14.4}{rm}$a$}}}
\put(311,660){\makebox(0,0)[lb]{\smash{\SetFigFont{12}{14.4}{rm}$d$}}}
\end{picture}

\end{center}

The Boltzmann weights satisfy the following relations.

\begin{itemize}
\item Yang-Baxter equation

\bac
&&
\sum_g W\left(\Matrix fgab{u-v}\right)
W\left(\Matrix gdbcu \right)
W\left(\Matrix fegdv \right) \\
= 
&& \sum_g W\left(\Matrix agbcv \right)
W\left(\Matrix feagu \right)
W\left(\Matrix edgc{u-v} \right).\label{yb}
\ea

\item Unitarity relation

\be
\sum_g W\left(\Matrix dgabu \right)
W\left(\Matrix dcgb{-u} \right)
=\delta_{ac}.\label{un}
\ee

\item Crossing relation

\be
W\left(\Matrix dcabu \right)
=\frac{[a]}{[b]}
W\left(\Matrix adbc{1-u} \right).\label{cr}
\ee

\end{itemize}
\subsection{The vertex operators}
Consider the corner transfer matrices
$A(\zeta), B(\zeta), C(\zeta), D(\zeta)$
which represent the $NW, SW, SE, NE$ quadrants,
respectively. 
Let $\CL_{l,k}$ be the space of the eigenvectors of
$A(\zeta)$ in the sector such that the central height is
equal to $k$ and the boundary heights are
$(l,l+1)$.
We denote by $\Phi^{(k+\varepsilon,k)}(\zeta)$
the half-infinite transfer matrix extending to infinity in the north.
This is an operator
\[
\Phi^{(k+\varepsilon,k)}(\zeta): \CL_{l,k}
\longrightarrow \CL_{l,k+\varepsilon}.
\]
Similarly, we denote by
$\Phi^{*(k,k+\varepsilon)}(\zeta^{-1})$
the half-infinite transfer matrix extending to
infinity in the west. This is an operator
\[
\Phi^{*(k,k+\varepsilon)}(\zeta^{-1}): \CL_{l,k+\varepsilon}
\longrightarrow \CL_{l,k}.
\]
They satisfy the following relations.
\begin{itemize}
\item Exchange relation
\be
\sum_g W\left(\Matrix agbc{u_2-u_1}\right)
\Phi^{(a,g)}(\zeta_2)\Phi^{(g,c)}(\zeta_1)
= \Phi^{(a,b)}(\zeta_1)\Phi^{(b,c)}(\zeta_2).\label{Iexchange}
\ee

\item Duality
\be
\Phi^{*(k,k+\varepsilon)}(\zeta)=\frac{[1]}{[k]}
\Phi^{(k,k+\varepsilon)}(x^2\zeta).\label{Iduality}
\ee

\item Inversion relation

\be
\sum_g \Phi^{*(a,g)}(\zeta)\Phi^{(g,a)}(\zeta) = 1, \quad
       \Phi^{(a,b)}(\zeta)\Phi^{*(b,c)}(\zeta) = \delta_{ac}.
\label{Iinversion}
\ee
\end{itemize}

\noindent
These operators are realized as  vertex operators acting 
on the bosonic Fock space ${\cal F}_{l,k}$
in Section 3.1.
\subsection{The boundary weights.}\label{sub1}
We follow Sklyanin's scheme in dealing with the boundary
ABF model \cite{Skl87}. Boundary weights 
\[
K\mmatrix k{k+\varepsilon}{k+\varepsilon'}u
\quad (\varepsilon,\varepsilon' = \pm)
\]
are given to the boundary
configurations. We restricts to the diagonal case, $\varepsilon=\varepsilon'$,
and use a solution of the boundary Yang-Baxter equation,
\bac
& & \sum_{f,g}
W\left( \Matrix cfba{u-v}\right)
W\left( \Matrix cdfg{u+v}\right)
K\mmatrix fgau
K\mmatrix degv \\
&=& \sum_{f,g}
W\left( \Matrix cdfe{u-v}\right)
W\left( \Matrix cfbg{u+v}\right)
K\mmatrix fegu
K\mmatrix bgav.
\ea
The solution is given by \cite{BPO95}
\be
\frac{K\mmatrix{k+1}kku}
     {K\mmatrix{k-1}kku}
=\frac{[c+u][k+c-u]}{[c-u][k+c+u]}.
\ee
Here the constant $c$ is arbitrary.
We choose to restrict $c$ to lie in one of two separate regions
parameterised by
A) $x^{2c}=-x^{2b}$,  $-1<b<1$; B) $x^{2c}=x^{2b}$,  $-1<b<1$.
We further restrict $u$ to satisfy  $0 < u < |b| < 1$ in both regions.
The overall factor in $K\mmatrix{k \pm1}kku$
is determined later.

Graphically, we represent 
$K\mmatrix abcu $ by
\begin{center}
\begin{picture}(0,0)%
\includegraphics{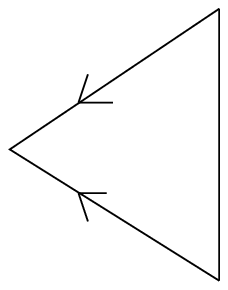}%
\end{picture}%
\setlength{\unitlength}{0.01250000in}%
\begingroup\makeatletter\ifx\SetFigFont\undefined
\def\x#1#2#3#4#5#6#7\relax{\def\x{#1#2#3#4#5#6}}%
\expandafter\x\fmtname xxxxxx\relax \def\y{splain}%
\ifx\x\y   
\gdef\SetFigFont#1#2#3{%
  \ifnum #1<17\tiny\else \ifnum #1<20\small\else
  \ifnum #1<24\normalsize\else \ifnum #1<29\large\else
  \ifnum #1<34\Large\else \ifnum #1<41\LARGE\else
     \huge\fi\fi\fi\fi\fi\fi
  \csname #3\endcsname}%
\else
\gdef\SetFigFont#1#2#3{\begingroup
  \count@#1\relax \ifnum 25<\count@\count@25\fi
  \def\x{\endgroup\@setsize\SetFigFont{#2pt}}%
  \expandafter\x
    \csname \romannumeral\the\count@ pt\expandafter\endcsname
    \csname @\romannumeral\the\count@ pt\endcsname
  \csname #3\endcsname}%
\fi
\fi\endgroup
\begin{picture}(98,126)(347,593)
\put(415,653){\makebox(0,0)[lb]{\smash{\SetFigFont{12}{14.4}{rm}$u$}}}
\put(445,595){\makebox(0,0)[lb]{\smash{\SetFigFont{12}{14.4}{rm}$c$}}}
\put(445,710){\makebox(0,0)[lb]{\smash{\SetFigFont{12}{14.4}{rm}$b$}}}
\put(347,653){\makebox(0,0)[lb]{\smash{\SetFigFont{12}{14.4}{rm}$a$}}}
\end{picture}

\end{center}
\subsection{The boundary transfer matrix.}
We define the boundary transfer matrix
\be
T_B^{(k)}(u)=\sum_{\varepsilon}\Phi^{*(k,k+\varepsilon)}(\zeta^{-1})
K\mmatrix {k+\varepsilon}kku
\Phi^{(k+\varepsilon,k)}(\zeta).\label{trans}
\ee
The boundary Yang-Baxter equation implies \cite{Skl87,BPO95}
\be
[T_B^{(k)}(u),T_B^{(k)}(v)]=0.
\ee
Graphically, $T_B^{(k)}(u)$ is represented by the following
half-infinite transfer matrix
\begin{center}
\begin{picture}(0,0)%
\includegraphics{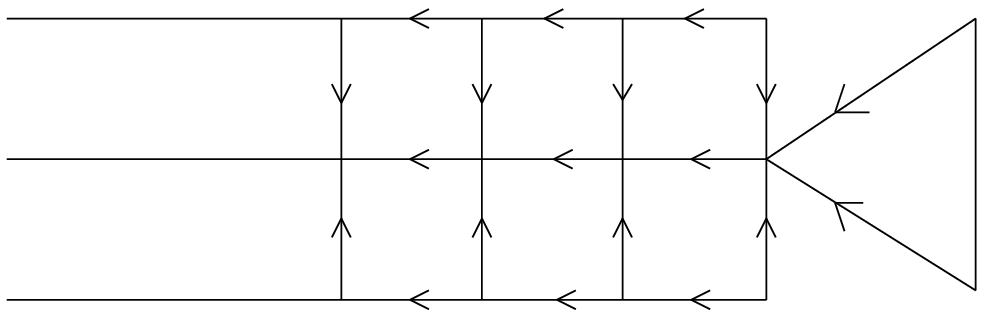}%
\end{picture}%
\setlength{\unitlength}{0.01250000in}%
\begingroup\makeatletter\ifx\SetFigFont\undefined
\def\x#1#2#3#4#5#6#7\relax{\def\x{#1#2#3#4#5#6}}%
\expandafter\x\fmtname xxxxxx\relax \def\y{splain}%
\ifx\x\y   
\gdef\SetFigFont#1#2#3{%
  \ifnum #1<17\tiny\else \ifnum #1<20\small\else
  \ifnum #1<24\normalsize\else \ifnum #1<29\large\else
  \ifnum #1<34\Large\else \ifnum #1<41\LARGE\else
     \huge\fi\fi\fi\fi\fi\fi
  \csname #3\endcsname}%
\else
\gdef\SetFigFont#1#2#3{\begingroup
  \count@#1\relax \ifnum 25<\count@\count@25\fi
  \def\x{\endgroup\@setsize\SetFigFont{#2pt}}%
  \expandafter\x
    \csname \romannumeral\the\count@ pt\expandafter\endcsname
    \csname @\romannumeral\the\count@ pt\endcsname
  \csname #3\endcsname}%
\fi
\fi\endgroup
\begin{picture}(316,136)(129,583)
\put(445,590){\makebox(0,0)[lb]{\smash{\SetFigFont{12}{14.4}{rm}$k$}}}
\put(445,705){\makebox(0,0)[lb]{\smash{\SetFigFont{12}{14.4}{rm}$k$}}}
\put(375,585){\makebox(0,0)[lb]{\smash{\SetFigFont{12}{14.4}{rm}$k$}}}
\put(375,710){\makebox(0,0)[lb]{\smash{\SetFigFont{12}{14.4}{rm}$k$}}}
\put(300,673){\makebox(0,0)[lb]{\smash{\SetFigFont{12}{14.4}{rm}$u$}}}
\put(345,673){\makebox(0,0)[lb]{\smash{\SetFigFont{12}{14.4}{rm}$u$}}}
\put(254,673){\makebox(0,0)[lb]{\smash{\SetFigFont{12}{14.4}{rm}$u$}}}
\put(254,627){\makebox(0,0)[lb]{\smash{\SetFigFont{12}{14.4}{rm}$u$}}}
\put(300,627){\makebox(0,0)[lb]{\smash{\SetFigFont{12}{14.4}{rm}$u$}}}
\put(345,627){\makebox(0,0)[lb]{\smash{\SetFigFont{12}{14.4}{rm}$u$}}}
\put(171,627){\makebox(0,0)[lb]{\smash{\SetFigFont{12}{14.4}{rm}$\cdots$}}}
\put(171,673){\makebox(0,0)[lb]{\smash{\SetFigFont{12}{14.4}{rm}$\cdots$}}}
\put(415,653){\makebox(0,0)[lb]{\smash{\SetFigFont{12}{14.4}{rm}$u$}}}
\end{picture}

\end{center}
We also define the boundary Hamiltonian $H_B^{(k)}$ in terms of the 
transfer matrix as 
\be H_B^{(k)}={(1-x^2)\ov 2x}   {\z d T_B^{(k)}(\z) \ov d \z}
{\big|}_{\z=1}.
\label{hamilton}\ee

Our first question is about
the ground state configuration, i.e., the largest eigenvalue 
eigenvector of 
$T_B^{(k)}(u)$ in the limit $x \to 0$.
We know that in the bulk theory the ground state
configuration of heights is given by the alternating sequence of
$l$ and $l+1$.
Because we have chosen to fix the heights at the right-most corners
to be equal to $k$, it follows that the possible choices  
are $l=k$ or $k-1$.
Let us compare $K\mmatrix {k\pm 1}kku$.
In fact, in the limit $x \to 0$ we have
\be
\frac{K\mmatrix{k+1}kku}
{K\mmatrix{k-1}kku}= %
\quad\left\{ \begin{array}{ll}
           >1 \mbox{\quad if \quad $b>0$;} \\
           <1 \mbox{\quad if \quad $b<0$.}
        \end{array}
\right.
\ee
This is true in both regions A and B.
Therefore, the ground state configuration for $b>0$
is given by $l=k$, and for $b<0$ it is given by 
$l=k-1$.

Next, let us fix the overall factor 
in the boundary weights. The normalisation will differ depending 
on whether $b\gl 0$, and we shall use the corresponding notation
$\nb{c}{\gl}{+\ep}{u}$ for the normalised boundary weights.

First, suppose $b>0$, and set
\begin{eqnarray*}
&&\bnb{c}{>}{+1}u=1,\\
&&\bnb{c}{>}{-1}u=
{[c-u][k+c+u]\over [c+u][k+c-u]}.
\end{eqnarray*}
The corresponding boundary transfer matrix
$\bar T^{(k)}_B(u)$ satisfies
\begin{eqnarray}
&&
\bar T^{(k)}_B(-u)
\bar T^{(k)}_B(u)=1,
\hskip8.1truecm
\\
&&
\bar T^{(k)}_B(u)=
{[2-2u][c+u][k+c-u]\over R(2u-1)[2u][1+c-u][k+c-1+u]}
\bar T^{(k)}_B(1-u).
\end{eqnarray}
Since $\{\bar T^{(k)}_B(u)\}$
is a commuting family, the eigenvalues of
$\bar T^{(k)}_B(u)$
also satisfy these equations. Assuming the analyticity
of $\log\bar t^{(k)}_B(u)$ for the largest eigenvalue $\bar t^{(k)}_B(u)$
in a neighbourhood of the annulus
$x\leq |\zeta|\leq 1$, we determine
$\bar t^{(k)}_B(u)$.
After that we choose the normalisation of
$\nb{c}{>}{\pm 1}u$
in such a way that 
$T^{(k)}_B(u)$
has the largest eigenvalue~1.
The result is as follows
\begin{eqnarray}
 \nb{c}{>}{+1}u &=& h^{(k)}_>(u),\label{Kp0}\\
 \nb{c}{>}{-1}u &=& h^{(k)}_>(u)
{[c-u][k+c+u]\over [c+u][k+c-u]},\label{Kp1}
\end{eqnarray}
\begin{eqnarray*}
 h^{(k)}_>(u) &=& \zeta^{r-1-2k\over 2r}
{f(\zeta)p^{(k)}_>(\zeta)p^{(k)}_>(x^2\zeta^{-1})\over
f(\zeta^{-1})p^{(k)}_>(\zeta^{-1})p^{(k)}_>(x^2\zeta)},\\
f(\zeta) &=&
{(x^{2r}\zeta^2;x^8,x^{2r})_\infty
(x^{8}\zeta^2;x^8,x^{2r})_\infty\over
(x^{6}\zeta^2;x^8,x^{2r})_\infty
(x^{2+2r}\zeta^2;x^8,x^{2r})_\infty},\\
p^{(k)}_>(\zeta) &=&
{(x^{2(1+c)}\zeta;x^4,x^{2r})_\infty
(x^{2(r-c-k+1)}\zeta;x^4,x^{2r})_\infty\over
(x^{2(r-c)}\zeta;x^4,x^{2r})_\infty
(x^{2(c+k)}\zeta;x^4,x^{2r})_\infty}.
\end{eqnarray*}

This normalisation of $\nb{c}{>}{\pm1}u$ is valid for
both regions $A$ and $B$ when $b>0$.
The case $b<0$ is similar. It is even not
necessary to repeat the calculation because we have a symmetry
$
(c, k,\pm)\rightarrow (-c, r-k,\mp).
$
Therefore, we have 
\begin{eqnarray}
\nb{c}{<}{+1}u &=&
h_<^{(k)}(u)
{\frac{[c+u][k+c-u]}{[c-u][k+c+u]}},\label{Km0}\\
\nb{c}{<}{-1}u &=&
h_<^{(k)}(u),\label{Km1}\eea
\begin{eqnarray*}
h_<^{(k)}(u)&=&
\zeta^{\frac{2k-1-r}{2r}}
{\frac{f(\zeta)p_<^{(k)}(\zeta)p_<^{(k)}(x^2\zeta^{-1})}
{f(\zeta^{-1})p_<^{(k)}(\zeta^{-1})p_<^{(k)}(x^2\zeta)}},\\
p_<^{(k)}(\zeta)&=&
{\frac{(x^{2(1-c)}\zeta; x^4, x^{2r})_\infty
(x^{2(c+k+1)};x^4, x^{2r})_\infty}
{(x^{2(r+c)}\zeta;x^4,x^{2r})_\infty
(x^{2(r-c-k)}\zeta; x^4,x^{2r})_\infty}}.
\end{eqnarray*}
In conclusion, we have fixed the normalisation of
the boundary weights for $b>0$ and $b<0$ 
separately so that the largest eigenvalue of the
boundary transfer matrix is 1.

For completeness, we give two relations for
$T_B^{(k)}(u)$ and the related relations for
$\nb{c}{\gl}{+\ep}u.$
\begin{eqnarray}
&&
T_B^{(k)}(-u)T_B^{(k)}(u)=1,\\
&&
T_B^{(k)}(1-u)=T_B^{(k)}(u),\\
&&
\nbp{c}{\gl}{k'}u \nbp{c}{\gl}{k'}{-u} =1,\\
&&
\nbp{c}{\gl}{k'}{1-u}=
\sum_{k''}W\left( \Matrix{k}{k'}{k''}{k}{2u-1}\right)
{[k'']\over[k']}\nbp{c}{\gl}{k''}u .
\end{eqnarray}
\setcounter{section}{2}
\setcounter{subsection}{2}
\section{Bosonization}
We follow \cite{LP96} to obtain a bosonic realization of
the spaces ${\cal L}_{l,k}$, their duals,
and the vertex operators.

\subsection{Boson Fock spaces}
We use the bosonic oscillators
\bea
[\b_m,\b_n]&=&m{[m]_x[(r-1)m]_x\ov[2m]_x[rm]_x}\d_{m+n,0},\\
{[}P,Q]&=&-i,
\ena
where $[k]_x={x^k-x^{-k}\over x-x^{-1}}$.
The Fock spaces are
\bea
{\cal F}_{l,k}&=&{\bf C}[\b_{-1},\b_{-2},\cdots]|l,k\rangle,\\
{\cal F}^*_{l,k}&=&\langle l,k|{\bf C}[\b_1,\b_2,\cdots],\\
\langle l,k|l,k\rangle&=&1,
\ena
where
\bea
\b_m|l,k\rangle&=&0\hbox{ if $m\ge1$,}\\
\langle l,k|\b_m&=&0\hbox{ if $m\leq-1$.}
\ena
The operator $P$ acts as
\bea
l\sqrt{r \ov 2(r-1)} -k \sqrt{r-1\ov 2r}
\ws{\rm on}\ws \cF_{l,k}\ws{\rm and}\ws \cF^*_{l,k}.
\ena
We also use operators $K$ and $L$ which act on
$\cF_{l,k}$, $\cF^*_{l,k}$ as $k$ and $l$, respectively.
We use the convention of left action on $\cF_{l,k}$
and right action on $\cF^*_{l,k}$. Namely, the composition
$AB$ of operators $A$ and $B$ is such that $B$ acts first
on $\cF_{l,k}$ and $A$ acts first on $\cF^*_{l,k}$.
\subsection{Type I and II vertex operators}
We define the type I vertex operators
\bea
\vo_+(\z)&=&\z^{-{r-1 \ov 4r}}:e^{-i \sqrt{r-1\ov 2r}(Q+iP\log\z)
-\sl_{m\ne 0}{\b_m\ov m}\z^{m}}:,\\[2mm]
\vo_-(\z)
&=&\l^{-1} \oint_{C}\dz\;
\vo_+(\z)A(z){[u+v+\half-K]\ov[u+v-\half]},\\[2mm]
A(z)&=&z^{r-1\ov r}:e^{i\sqrt{2(r-1)\ov r}(Q-iP\log z)
+\sl_{m\ne 0}{\b_m\ov m} (x^m+x^{-m})z^{-m}}:,
\ena
where
\bea
&&\z=x^{2u},\quad z=x^{2v},\\
&&\l=x^{{1-r\ov 2r}} \qp{x^{2r-2}}{x^{2r}} { \mcb{x^4}\mcb{x^{4+2r}} \ov
\mcb{x^2}\mcb{x^{2+2r}} }.
\ena
The contour $C$ for the $z$-integration in $\Phi_-(\z)$ is chosen in
such a way that the poles from the factor $[u+v-\half]$ at $z=x^{1+2nr}/\z$
are inside if $n\in\Z_{\geq0}$ (outside if $n\in\Z_{<0})$),
and the poles from the normal ordering of the product
$\vo_+(\z)A(z)$ at $z=1/x^{1+2nr}\z$ $(n\in\Z_{\geq0})$
are outside. The operator $\Phi_\pm(\z)$
acts as
\bea
\Phi_\pm(\zeta)&:&\cF_{l,k}\rightarrow\cF_{l,k\pm1},\\
\Phi_\pm(\zeta)&:&\cF^*_{l,k\pm1}\rightarrow\cF^*_{l,k}.
\ena
The value of $K$ in $\Phi_-(\zeta)$ is $k$ in both cases.

The type II vertex operators are
\bea
\vot_+(\z) &=&
\z^{-{r\ov 4(r-1)}}:e^{i\sqrt{r\ov 2(r-1)}(Q+iP\log\z)
+ \sl_{m\ne 0}  {\a_m\ov m}\z^{m}}:,\\[3mm]
\vot_-(\z)&=& \oint_{C'}\dz\;
\vot_+(\z)B(z){[\half-L-u-v]^{\pr}\ov[u+v+\half]^{\pr}},\\[3mm]
\ena
where
\bea
\a_m&=&(-1)^m{[rm]_x\ov[(r-1)m]_x}\b_m,\\
B(z)&=&
z^{r\ov r-1}:e^{-i\sqrt{2r\ov r-1}(Q-iP\log z)
-\sl_{m\ne 0}{\a_m\ov m} (x^m+x^{-m})z^{-m}}:,\\[3mm]
[u]'&=&x^{{u^2\over r-1}-u}\Theta_{x^{2(r-1)}}(x^{2u}).
\ena
The contour $C'$ for the $z$-integration in $\Psi^*_-(\z)$ is chosen in
such a way that the poles from the factor $[u+v+\half]'$ at
$z=x^{2n(r-1)-1}/\z$
are inside if $n\in\Z_{\geq0}$ (outside if $n\in\Z_{<0}$)
and the poles from the normal ordering of the product
$\vot_+(\z)B(z)$ at $z=x^{1-2n(r-1)}/\z$ $(n\in\Z_{\geq0})$ are outside.
The operator $\vot_\pm(\z)$ acts as
\bea
\vot_\pm(\zeta)&:&\cF_{l,k}\rightarrow\cF_{l\pm1,k},\\
\vot_\pm(\zeta)&:&\cF^*_{l\pm1,k}\rightarrow\cF^*_{l,k}.
\ena
The value of $L$ in $\Psi^*_-(\zeta)$ is $l$ in both cases.

The vertex operators satisfy the following commutation
relations.
\bea
&&\Phi_{\varepsilon_2}(\zeta_2)\Phi_{\varepsilon_1}(\zeta_1)
\nonumber\\[2mm]
&&=\sum_{\varepsilon'_1+\varepsilon'_2=
\varepsilon_1+\varepsilon_2}
W\left(\Matrix
{K+\varepsilon_1+\varepsilon_2}
{K+\varepsilon'_2}
{K+\varepsilon_1}
K
{u_1-u_2}\right)\Phi_{\varepsilon'_1}(\zeta_1)\Phi_{\varepsilon'_2}(\zeta_2)
\label{exch},\\[3mm]
&&\phantom{mmm}\sl_{\varepsilon}\vo_{\varepsilon}^*(\z)
\vo_{\varepsilon}(\z)=1,\label{unit}\\[2mm]
&&\phantom{mmm}\vo_{\varepsilon'}(\z)\vo_{\ep}^*(\z)=\d_{\varepsilon'\ep},
\label{inver}
\ena
where
\be
\vo_{\varepsilon}^*(\z)={[1]\over[K]}
\vo_{-\varepsilon}(x^2\z).
\label{Idual}\ee
Comparing \mref{exch}-\mref{inver} with
\mref{Iexchange}-\mref{Iinversion},
we see that $\vo_{\ep}(\z)$ and $\vo_{\ep}^*(\zi)$ provide a Fock space
realisation of the half-infinite transfer matrices
$\vo^{(k+\ep,k)}(\z)$
and $\vo^{*\,(k,k+\ep)}(\zi)$ respectively.

We also have
\bea
&&\vot_{\varepsilon_2}(\z_2) \vot_{\varepsilon_1}(\z_1)\nonumber\\
&&=-\sl_{\varepsilon'_1+\varepsilon'_2=\varepsilon_1+\varepsilon_2}
\newwet{L+\varepsilon_1+\varepsilon_2}{L+\varepsilon'_2}
{L+ \varepsilon_1}{L}{u_2-u_1}
\vot_{\varepsilon'_1}(\z_1) \vot_{\varepsilon'_2}(\z_2),\label{comm1}
\nonumber\\[3mm]\\
&&\vo_{\varepsilon_1}(\z_1) \vot_{\varepsilon_2}(\z_2)
=\chi( {\z_2 / \z_1})\vot_{\varepsilon_2}(\z_2)
\vo_{\varepsilon_1}(\z_1),
\label{comm2}
\ena
where $\newwet dcabu$
is the Boltzmann weight with $r$ replaced by $r-1$, and
\bea
\chi(\z)=\z^{-1/2}{\Theta_{x^4}(-x\z)\ov\Theta_{x^4}(-x\zi)}.
\ena

Set
\be
\Psi_\varepsilon(\z)=\Psi^*_{-\varepsilon}(x^2\z){[1]'\over[L]'}.
\label{IIdual}\ee
We then have a pole
\be
\Psi_\varepsilon(\z_1)\Psi^*_\varepsilon(\z_2)=
{\varepsilon g\over1-\z_1/\z_2}+\cdots,\label{IIpole}
\ee
where
\bea
g&=&
{x^{r\over2(r-1)}}
{(x^{2r};x^{2(r-1)})_\infty
(x^2;x^4,x^{2(r-1)})_\infty
(x^{2r};x^4,x^{2(r-1)})_\infty
\over
(x^{2(r-1)};x^{2(r-1)})_\infty
(x^4;x^4,x^{2(r-1)})_\infty
(x^{2(r-1)};x^4,x^{2(r-1)})_\infty}.
\ena

Our bosonization is slightly different from than that of reference
\cite{LP96}. Firstly, we have changed $\z$ to $\z^{-1}$ in the definition
of $\vo_{\pm}$. Secondly, our $\vo_-$ involves an extra factor
of $(-1)^{K-L}$ over the corresponding object $\Psi^-$ of 
\cite{LP96}. This factor arises because our Boltzmann weights
$W$ differ from $U$ of \cite{LP96} by a $-$ sign in \mref{w3}.
This difference is simply a gauge transformation.
Finally, our $\a_m$ is equal to $(-1)^m\a_m$ of \cite{LP96}. We introduce
this factor in order to give the correct sign in the commutation
relations \mref{comm2}.

\subsection{Screening operators}
We define two screening operators $X$ and $Y$ that have the properties
\cite{LP96,JLMP}.
\bea
&&X:\cF_{l,k}\rightarrow\cF_{l-2,k}\quad
(\cF^*_{l-2,k}\rightarrow\cF^*_{l,k}),\\
&&Y:\cF_{l,k}\rightarrow\cF_{l,k-2}\quad
(\cF^*_{l,k-2}\rightarrow\cF^*_{l,k}),\\
&&[\Phi_\varepsilon(\z),X]=0,\quad
[\Psi^*_\varepsilon(\z),Y]=0,\quad
[X,Y]=0,\label{3.60}\\
&&X^{r-1}=0,\quad Y^r=0.\label{3.61}
\ena
They have the following bosonization.
\bea
X&=&\oint{dz\over2\pi iz}B(z){[\half-L-v]'\over[\half+v]'},\\
Y&=&\oint{dz\over2\pi iz}A(z){[\half-K+v]\over[\half-v]}.\\
\ena
It is proved in \cite{JLMP} that these expressions obey
\mref{3.61}.

Fix $l$ $(1\leq l\leq r-1)$ and $k$ $(1\leq l\leq r)$.
We have the following complexes.

\medskip
\bea
\begin{diagram}
\node{\cdots}\arrow{e,t}{X_{-2}=X^l}
\node{\cF_{2(r-1)-l,k}}\arrow{e,t}{X_{-1}=X^{r-1-l}}
\node{\cF_{l,k}}\arrow{e,t}{X_0=X^l}
\node{\cF_{-l,k}}\arrow{e,t}{X_1=X^{r-1-l}}
\node{\cdots}
\end{diagram}
\ena

\medskip
\bea
\begin{diagram}
\node{\cdots}
\node{\cF^*_{2(r-1)-l,k}}\arrow{w,t}{X^*_{-2}=X^l}
\node{\cF^*_{l,k}}\arrow{w,t}{X^*_{-1}=X^{r-1-l}}
\node{\cF^*_{-l,k}}\arrow{w,t}{X^*_0=X^l}
\node{\cdots}\arrow{w,t}{X^*_1=X^{r-1-l}}
\end{diagram}
\ena

\medskip
\bea
\begin{diagram}
\node{\cdots}\arrow{e,t}{Y_{-2}=Y^k}
\node{\cF_{l,2r-k}}\arrow{e,t}{Y_{-1}=Y^{r-k}}
\node{\cF_{l,k}}\arrow{e,t}{Y_0=Y^k}
\node{\cF_{l,-k}}\arrow{e,t}{Y_1=Y^{r-k}}
\node{\cdots}
\end{diagram}
\ena

\medskip
\bea
\begin{diagram}
\node{\cdots}
\node{\cF^*_{l,2r-k}}\arrow{w,t}{Y^*_{-2}=Y^k}
\node{\cF^*_{l,k}}\arrow{w,t}{Y^*_{-1}=Y^{r-k}}
\node{\cF^*_{l,-k}}\arrow{w,t}{Y^*_0=Y^k}
\node{\cdots}\arrow{w,t}{Y^*_1=Y^{r-k}}
\end{diagram}
\ena

We assume the following cohomological properties.
\bea
{\rm Ker\,}X_j/{\rm Im\,}X_{j-1}&=&0\hbox{ if $j\not=0$},\\[2mm]
{\rm Ker\,}X^*_{j-1}/{\rm Im\,}X^*_j&=&0\hbox{ if $j\not=0$},\\[2mm]
{\rm Ker\,}X_0&=&{\rm Ker\,}Y_0,\\[2mm]
{\rm Im\,}X_{-1}&=&{\rm Im\,}Y_{-1},\\[2mm]
{\rm Ker\,}Y_j/{\rm Im\,}Y_{j-1}&=&0\hbox{ if $j\not=0$},\\[2mm]
{\rm Ker\,}Y^*_{j-1}/{\rm Im\,}Y^*_j&=&0\hbox{ if $j\not=0$},\\[2mm]
{\rm Ker\,}X^*_{-1}&=&{\rm Ker\,}Y^*_{-1},\\[2mm]
{\rm Im\,}X^*_0&=&{\rm Im\,}Y^*_0.
\ena
We will make the identification,
\bea
{\cal L}_{l,k}&=&{\rm Ker\,}X_0/{\rm Im\,}X_{-1},\\
{\cal L}^*_{l,k}&=&{\rm Ker\,}X^*_{-1}/{\rm Im\,}X^*_0,
\ena
and conjecture that the coupling
\bea
{\cal L}^*_{l,k}\times{\cal L}_{l,k}\rightarrow{\bf C},
\ena
induced from the coupling
\bea
{\cal F}^*_{l,k}\times{\cal F}_{l,k}\rightarrow{\bf C}
\ena
is non-degenerate.

In summary, we have bosonized the half-infinite transfer matrices
$\Phi^{(k\pm1,k)}(\z)$ and $\Phi^{*\,(k,k\pm1)}(\zi)$ by means of the type I vertex operators
$\Phi_\pm(\z)$ and $\vo^{*}_{\pm}(\zi)$, and also introduced type II vertex operators,
which will play the role of creation operators of particles
as we will see in the following section.
\setcounter{equation}{0}
\setcounter{section}{3}
\section{The Boundary States}

\ss{Construction in the Fock space}
In Section 2, we chose the normalisation of the
bulk and boundary weights such that
the maximal eigenvalues of the bulk and boundary transfer matrices
in the large lattice limit were $1$.
In this section we
describe how to construct the corresponding eigenvector in 
the bosonic Fock space $\cF_{k,k}$ or $\cF_{k-1,k}$, depending
on whether $b>0$ or $b<0$. We will then show that the eigenvector
so constructed actually determines a non-zero residue class
in the corresponding subquotient $\cL_{k,k}$ or $\cL_{k-1,k}$.

The maximal eigenvectors $\rbs{k}$ and $\rbs{k-1}$
are defined in both regions A and B (defined in Section \ref{sub1}) by,
\bea 
T_B^{(k)}(\z) \rbs{k} &=& \rbs{k}\in \cF_{k,k}, \quad\quad 
{\rm for}\ws b>0,\label{cond0}\\[2mm]
T_B^{(k)}(\z) \rbs{k-1} &=& \rbs{k-1}\in \cF_{k-1,k} , \quad\quad {\rm for}\ws b<0.\label{cond1}
\eea
We make the Ansatz that
\be
\rbs{k-i} = e^{F_i^{c,k}} \ket{k-i,k}, {\rm for}\ws i=0,1,\label{ansatz}
\ee
where 
\be
F_i^{c,k}= -{\half} \sl_{m>0} 
{[2m]_x [rm]_x \ov m [m]_x [(r-1)m]_x } \b_{-m}^2 +
\sl_{m>0}{D_{m,i}^{c,k} \ov m} \b_{-m}.
\ee
The adjoint action of $e^{F_i^{c,k}}$ on the bosonic oscillator modes is that
of a Bogoliubov transformation. Namely,
\bac
e^{-F_i^{c,k}} \b_{m} e^{F_i^{c,k}} &=& \b_m - \b_{-m} + 
{[m]_x[(r-1)m]_x \ov [2m]_x [rm]_x} D_{m,i}^{c,k}, \\[2mm]
e^{-F_i^{c,k}} \b_{-m} e^{F_i^{c,k}} &=& \b_{-m},
\ea
where $m>0$.
The coefficients $D_{m,i}^{c,k}$ are determined by solving conditions
(\ref{cond0}) and (\ref{cond1}). Rewriting these conditions in
terms of the bosonic vertex operator expression for 
$T_B^{(k)}(\z)$ (\ref{trans}), and using
the inversion property (\ref{inver}), we obtain
\bea
\nb{c}{>}{+\ep}{u} \vo_{\ep}(\z) e^{F_0^{c,k}} \ket{k,k} =
 \vo_{\ep}(\zi) e^{F_0^{c,k}} \ket{k,k} 
\quad{\rm for}\ws b>0,\label{bsc1}\\[2mm]
\nb{c}{<}{+\ep}{u} \vo_{\ep}(\z) e^{F_1^{c,k}} \ket{k-1,k} =
 \vo_{\ep}(\zi) e^{F_1^{c,k}} \ket{k-1,k}
\quad{\rm for}\ws b<0.\label{bsc2}
\eea
Solving these conditions for $\ep=+$, we find
\bea
D_{m,0}^{c,k} &=&
- {[(k-1)m]_x \bpl{(r-2c-k)m} \ov [(r-1)m]_x} 
- \t_m\left( 
{[m/2]_x \bpl{rm/2} \ov [(r-1)m/2]_x}
\right),\label{D0def}\\[2mm]
D_{m,1}^{c,k} &=&
 {[r+1-k)m]_x \bpl{(2c+k)m} \ov [(r-1)m]_x} 
- \t_m\left( 
{[m/2]_x \bpl{rm/2} \ov [(r-1)m/2]_x}
\right)\label{D1def}\\[2mm]
&=& D_{m,0}^{c,k} + {[rm]_x \bpl{(2c+1)m} \ov [(r-1)m]_x},
\eea
where
\ba{llll}
[k]^+_x&=&x^k+x^{-k},\\[2mm]
\theta_m(x)&=&\cases{x& if $m$ is even;\cr
0& otherwise.\cr}
\ea
In order to show that (\ref{bsc1}) and (\ref{bsc2}) hold for
$\ep=-$, with $D_{m,i}^{c,k}$ given by (\ref{D0def}) and (\ref{D1def}),
we first consider (\ref{bsc1}) 
for $k=1$. We find that,
\bac
&&e^{-F_0^{c,1}} \vo_-(\z)e^{F_0^{c,1}}  |1,1\rangle = \l^{-1} \z^{{3r-1 \ov 4r}}f(\z)^{-1} \\
&&\quad \times \oint \dz
(z-z^{-1}) {\br{x^{2r-1}\z z} \br{x^{2r-1}\z/ z} \ov \br{x\z z}\br{x
\z/z}}
e^{\sl_{m<0} {\b_m\ov m} \left( (\z^m+\z^{-m}) -(\x^m+\x^{-m})
(z^m+z^{-m})
\right)} |1,1\rangle =0. 
\ea
The integral vanishes because of the anti-symmetry of the integrand under
the change of variables $z\ra z^{-1}$.
The proof that (\ref{bsc1}) and (\ref{bsc2}) 
are valid for $\ep=-$ and general $k$ is inductive
and is given in Appendix \ref{ap2}. 


In a similar manner we construct dual maximal eigenstates in the Fock spaces
$\cF^{*}_{k,k}$ and $\cF^{*}_{k-1,k}$. These states are determined
by the requirements,
\bea
 \lbs{k}T_B^{(k)}(\z) &=& \lbs{k} \in \cF_{k,k}^{*}
\quad {\rm for}\ws b>0,\label{dcond0}\\[2mm]
 \lbs{k-1}T_B^{(k)}(\z) &=& \lbs{k-1} \in \cF_{k-1,k}^{*}\
\quad {\rm for}\ws b<0.\label{dcond1}
\eea
As above we make the Ansatz that
\be
\lbs{k-i} = \bra{k-i,k} e^{G^{c,k}_i},
\ee
where
\be
G^{c,k}_i=-{\half} \sl_{m>0} 
x^{4m}{[2m]_x [rm]_x \ov m [m]_x [(r-1)m]_x } \b_{m}^2 +
\sl_{m>0}{E_{m,i}^{c,k} \ov m} \b_{m}.
\ee
$e^{G^{c,k}_i}$ now produces the Bogoliubov transformation
\bea
e^{G^{c,k}_i} \b_{-m} e^{-G^{c,k}_i} &=& \b_{-m}-x^{4m}\b_m +
{[m][(r-1)m] \ov [2m][rm]}  E_{m,i}^{c,k},\\[2mm]
e^{G^{c,k}_i} \b_{m} e^{-G^{c,k}_i} &=& \b_m,
\eea
where $m>0$. In terms of bosonized vertex operators, (\ref{dcond0}) and
(\ref{dcond1}) become
\bea
\lbs{k}\vo_\ep^{*} (\zi)\nb{c}{>}{+\ep}{u} =
\lbs{k} \vo_\ep^{*}(\z)
\quad {\rm for}\ws b>0,\\[2mm]
\lbs{k-1}\vo_\ep^{*} (\zi)\nb{c}{<}{+\ep}{u} =
\lbs{k-1} \vo_\ep^{*}(\z)
\quad {\rm for}\ws b<0.
\eea
Solving these equation, we find
\bea 
E_{m,0}^{c,k} &=& 
-x^{2m} { [(k+1)m]\bpl{(r-2c-k)m} \ov [(r-1)m]}
+x^{2m} \t_m\left( 
{[m/2]_x \bpl{rm/2} \ov [(r-1)m/2]_x} \right), \\[2mm]
E_{m,1}^{c,k} &=& 
x^{2m} { [(r-k-1)m]\bpl{(2c+k)m} \ov [(r-1)m]}
+x^{2m} \t_m\left( 
{[m/2]_x \bpl{rm/2} \ov [(r-1)m/2]_x} \right)\\[3mm]
&=& E_{m,0}^{c,k} +x^{2m} {[rm]_x \bpl{(1-2c)m} \ov [(r-1)m]} .
\eea

\ss{Boundary vacuum states in the subquotients}
We have constructed maximal eigenvectors $\rbs{k-i,k} \in
\cF_{k-i,k}$, and dual eigenvectors  $\lbs{k-i,k} \in \cF^{*}_{k-i,k}$.
In order to establish that these vectors actually
give rise to the eigenvectors in the
subquotients $\cL_{k-i,k}=\Ker X_0/\Im X_{-1}$ and 
$\cL^{*}_{k-i,k}=\Ker X^{*}_{-1}/\Im X_{0}^{*}$ respectively, 
it is sufficient to show that,
\bea
\rbs{k-i} &\in& \Ker X_0=\Ker Y_0\subset \cF_{k-i,k},\label{con1}\\
\lbs{k-i} &\in& \Ker X^{*}_{-1}=\Ker Y^{*}_{-1}\subset
\cF^{*}_{k-i,k},\label{con2}\\
\lbs{k-i}k-i,k\rangle_B^c &\neq& 0\label{con3}.
\eea
If these three conditions are true then it is simple to show 
that 
\be
\lbs{k-i}P\rbs{k-i}= \lbsb{k-i}P\rbsb{k-i},
\label{mat1}\ee
where $\rbsb{k-i}$ is the residue class of $\rbs{k-i}$ in $\cL_{k-i,k}$,\ \
$\lbsb{k-i}$ is the residue class of $\lbs{k-i}$ in $\cL^{*}_{k-i,k}$,
and $P$ is an operator that commutes with the coboundary operators $X$
and $Y$.
Local operators of the of the theory
are expressed in terms of products of vertex operators
$\vo_{\varepsilon}(z)$, and thus an example of an operator $P$.
Hence if (\ref{con1}) - (\ref{con2}) hold 
we can calculate local correlation functions of the theory as matrix
elements of the form  $\lbs{k-i}P\rbs{k-i}$. Form factors are also
in this class because they are the matrix elements of products
of type I and type II vertex operators which commute with $X$ and
$Y$ respectively - see (\ref{3.60}).

We establish condition (\ref{con1}) as follows. For
$\ket{1,1}_B^c=e^{F_0^{c,1}} \ket{1,1}$, one
can show the result directly. Namely,
\be 
e^{-F^{c,1}_0} X e^{F^{c,1}_0} \ket{1,1} \propto \oint \dz 
(z-z^{-1}) 
e^{\slp {\a_{-m}\ov m} {[m]\ov [2m]} (z^m+z^{-m})}
\ket{-1,1}=0.
\ee
This integral is zero because of the anti-symmetry of the integrand under
$z\ra z^{-1}$. For general $\rbs{k-i}$ this technique is rather complicated
and it is simpler to proceed by other means. We observe that different
boundary states are related to each other by the action of vertex
operators. In particular we find that
\bea
\vo_+(x^{2c}) \rbs{k} &\sim & \ket{k,k+1}_B^{c-1},\label{voact1} \\
{\rm Res}_{\xi=-x^{1+2c}} \vot_+(\xi) \rbs{k-1} &\sim & \rbs{k}. 
\label{voact2}\eea
Here $\sim$ means equal up to scalar functions. We shall discuss
\mref{voact2} and other relations in more detail in Section 6. 
These relations
are useful in this context because the screening charge $X$ commutes
with $\vo(\xi)$, and $Y$ commutes with $\vot(\xi)$.
Thus 
\ba{lllll}
X \ket{1,2}_B^{c} &\sim& \vo_+(x^{2(c+1)}) X \ket{1,1}_B^{c+1} &=&0,\\
Y^2 \ket{2,2}_B^c &\sim& {\rm Res}_{\xi=-x^{1+2c}}\vot_+(\xi)
Y^2 \ket{1,2}_B^c &=&0,
\ea
where $\ket{1,1}_B^{c+1}$ is an analytic continuation in $c$ of
$\ket{1,1}_B^{c}$. By repeatedly applying vertex operators to
construct the general state  $\rbs{k-i}$, we can show that
$\rbs{k-i}\in \Ker X_0=\Ker Y_0$ as required.

We establish condition (\ref{con2}) in a completely analogous manner.
First we show the result for ${_{B}^{c}}\langle {r-2,r-1} |$ directly
using,
\be 
\langle {r-2,r-1} | e^{G_1^{c,r-1}} X e^{-G_1^{c,r-1}}
\propto
\langle {r,r-1} | \oint \dz (zx^2 - z^{-1} x^{-2}) 
e^{-\slp {x^{2m}\a_{m}\ov m} {[2m]\ov [m]} (z^m x^{2m}+z^{-m}x^{-2m})}=0.
\ee
This integral vanishes because of anti-symmetry under the changes of
variables $z x^2 \ra z^{-1}x^{-2}$. Again, different dual states
are related to each other under the action of vertex operators. We
find,
\bac 
\lbs{k-1} \vo_-^{*}(x^{2c}) &\sim& {_{B}^{c+1}}\langle k-1,k-1 |,\\
\Res_{\xi= -x^{1-2c}}\; \lbs{k} \Psi_-(\xii) &\sim& \lbs{k-1}.
\ea
Thus
\bac 
{_{B}^{c}}\langle r-2,r-2 |X &\sim& {_{B}^{c-1}}\langle r-2,r-1|X
\vo_-^{*}(x^{2(c-1)})  =0\\
{_{B}^{c}}\langle r-3,r-2 |Y^2 &\sim& {\rm Res}_{\xi=-x^{1-2c}}\,
 {_{B}^{c}}\langle r-2,r-2|Y^2 \Psi_-(\xii)  =0 \quad{\rm etc}.
\ea
The general result $\lbs{k-i} \in \Ker X_{-1}^{*}= \Ker Y_{-1}^{*}$
follows.



Finally, we show (\ref{con3}) by computing 
$\lbs{k-i}k-i,k\rangle_B^c$
explicitly. We do this by using a decomposition of the
identity on $\cF_{l,k}$ which employs coherent states.
Define the coherent states
\bac
\ket{\xi} &=& e^{\slp \xi_m \kappa_m \b_{-m}} \ket{l,k}, \quad
\bra{\xib} = \bra{l,k} \, e^{\slp \xib_m \kappa_m \b_{m}}, \quad \quad{\rm
where}, \\[3mm]
\kappa_m&=& {1\ov m} {[2m]_x [rm]_x \ov [m]_x [(r-1)m]_x}. 
\ea
It is simple to show that
\bac
\b_{m} \ket{\xi} &=& \xi_m  \ket{\xi},\quad
\bra{\xib} \b_{-m} = \bra{\xib} \xib_{m}, \\[2mm]
\bra{\xib}\xi\rangle &=& e^{\slp \kappa_m \xib_m \xi_m }. \ea
Its is possible to decompose the identity on $\cF_{l,k}$ as
\be {\rm id}_{\cF_{l,k}}=\int\limits_{-\infty}^{\infty} \plp d\xi_m d\xib_m \kappa_m 
e^{-\slp \kappa_m \xib_m \xi_m}
\ket{\xi}_{l,k}\;  {_{l,k}}\bra{\xib}.
\ee
This is shown using the Gaussian integral
\ba{ll}
 &\int\limits_{-\infty}^{\infty} \plp d\xi_m d\xib_m \kappa_m 
e^{\left(
-\half \slp \kappa_m (\xib_m,\xi_m) {\cal{A}}_m \left(\xib_m \atop
\xi_m\right)
+\slp (\xib_m,\xi_m) {\cal{B}}_m\right)} \\[3mm]
& = 
\plp (-{\rm det}(\cA_m))^{-\half} e^{(\half \slp \kappa_m^{-1} \cB_m^t \cA_m^{-1}
\cB_m)}. 
\label{gauss}\ea
Inserting this decomposition of the identity between $e^{G_i^{c,k}}
e^{F_i^{c,k}}$ and using \mref{gauss}, we find,
\bac \label{NORM}
&&\lbs{k-i}k-i,k\rangle_B^c=\langle k-i,k|e^{G_i^{c,k}} e^{F_i^{c,k}}|k-i,k\rangle\\[3mm]&&=
{1\ov \qp{x^4}{x^4}^{\half}} \exp\left(
-\half \sl_{m>0} {
x^{4m} (D_{m,i}^{c,k})^2 -2D_{m,i}^{c,k}E_{m,i}^{c,k} 
+(E_{m,i}^{c,k})^2 
\ov \kappa_m (1-x^{4m}) m^2}
\right),
\ea
For $i=0$ this is equal to
\bac
&&\hspace*{-15mm}\lbs{k}k,k\rangle_B^c = { 1 \ov \qp{x^6}{x^8}}
{
 \qp{x^8}{x^8,x^{2r}} \qp{x^4}{x^8,x^{2(r-1)}} 
  \ov
 \qp{x^8}{x^8,x^{2(r-1)}} \qp{x^4}{x^8,x^{2r}} 
} 
\\[3mm]
&&\hspace*{-15mm}\times
{
\qp{x^{4r-4c-2k}} {x^4,x^{2(r-1)}} \qp{x^{4c +2k}} {x^4,x^{2(r-1)}}
 (x^{2r};x^4,x^{2(r-1)})_{\infty}^2 
\qp{x^{2r+4-2k}}{x^4,x^{2r}} \qp{x^{2r+4+2k}}{x^4,x^{2r}} 
\ov 
\qp{x^{2+4r-4c-2k}} {x^4,x^{2r}} \qp{x^{2+4c +2k}} {x^4,x^{2r}}
 (x^{2+2r};x^4,x^{2r})_{\infty}^2 
\qp{x^{2r+2-2k}}{x^4,x^{2(r-1)}} \qp{x^{2r+2+2k}}{x^4,x^{2(r-1)}} 
}
\\[3mm]&&\hspace*{-15mm}\times
{
\qp{x^{4r+4-4k-4c}}{x^8;x^{2r}} \qp{x^{4+4c}}{x^8;x^{2r}} 
\qp{x^{4r+4-4c}}{x^8;x^{2r}} \qp{x^{4+4k+4c}}{x^8;x^{2r}} 
\ov
\qp{x^{4r+2-4k-4c}}{x^8;x^{2(r-1)}} \qp{x^{2+4c}}{x^8;x^{2(r-1)}} 
\qp{x^{4r+2-4c}}{x^8;x^{2(r-1)}} \qp{x^{2+4k+4c}}{x^8;x^{2(r-1)}} 
}.
\ea
From the explicit form of $D_{m,i}^{c,k}$ and $E_{m,i}^{c,k}$, we
see we can recover the $i=1$ normalisation
from the $i=0$ result by the transformation $(c,k)\ra (c+r,k-r)$.
\setcounter{equation}{0}
\setcounter{section}{4}
\section{General Eigenstates and The Boundary S-Matrix}

\ss{General Eigenstates of $T_B^{(k)}(\z)$}
We construct general eigenstates of $T_B^{(k)}(\z)$ by using type II vertex
operators. The technique closely parallels that used for the XXZ
model \cite{JM94,JKKKM}.
From (\ref{trans}) and (\ref{comm2}) it follows that
\bac
T_B^{(k)}(\z)\vot_{\ep}(\xi) &=&t(\z,\xi)\vot_{\ep}(\xi)T_B^{(k)}(\z), 
\quad{\rm where,}\\[3mm]
t(\z,\xi)&=&
 \chi(\xi/\z)
\chi(\xi\z/x^2)
= \chi(\xi/\z) 
\chi(1/(\xi \z)).
\ea
From the definition \mref{hamilton} of the Hamiltonian $H_B^{(k)}$ and the 
property $T_B^{(k)}(1)={\rm id}$ (which follows from the definition 
\mref{trans}), it also follows that,
\bac
{[}H_B^{(k)},\vot_{\ep}(\xi)] &=&\epsilon(\xi)\vot_{\ep}(\xi), 
\quad{\rm where,}\\[3mm]
\epsilon(\xi)&=&{2I \ov \pi} \sinh ({\pi I'\ov I}) {\rm dn} ({2I \th \ov \pi}),
\quad
\xi=-e^{2 i \th}.
\ea
Here, dn is the Jacobi elliptic function with nome $x$,
and $I$ and $I'$ are the associated half-period magnitudes \cite{Bax82}.

Thus it is possible to construct general continuous 
eigenstates of the form 
\ba{rlll}
\vot_{\ep_1}(\xi_1) \cdots \vot_{\ep_n}(\xi_n) 
\ket{k,k}_B^c  &\in& \cF_{k+\ep_1+\cdots+\ep_n,k}
 &\quad{\rm for} \ws b>0,  \\
\vot_{\ep_1}(\xi_1) \cdots \vot_{\ep_n}(\xi_n) 
\ket{k-1,k}_B^c  &\in& \cF_{k-1+\ep_1+\cdots+\ep_n,k}
&\quad{\rm for} \ws b<0, \label{cstates},\ea
with $|\xi_i|=1$. The physical picture
of these excitations is of $n$ kinks or domain walls that are
free, i.e., not bound to the right boundary.

\ss{The Boundary S-Matrix}
The boundary S-matrix describes the scattering of these kink
states off the right boundary.
Following \cite{GZ}, we define the boundary S-matrix
$M_{\ep}^{(k-i,k)} (\xi;r,c)$ by
\bea
\vot_{\ep}(\xi) \ket{k-i,k}_B^c &=& M_{\ep}^{(k-i,k)} (\a;r,c)
\vot_{\ep}(\xii)\ket{k-i,k}_B^c \quad {\rm for}\ws i=0,1,
\label{scond}\\[3mm]
\lbs{k-i} \Psi_{\ep}(\xii) &=& \lbs{k-i} \Psi_{\ep}(\xi) M_{\ep}^{(k-i,k)}
(\a;r,c),\label{dscond}
\eea
where $\xi=x^{2\a}$. 
Solving condition (\ref{scond}) and (\ref{dscond})
We find,
\bea
M_{\pm}^{(k,k)}(\a;r,c)&=&-\xi 
{ \t_{x^8}(\xi^{-2}x^2) \ov \t_{x^8} (\xi^2 x^2) }
\nbt{c-\half+\d}{>}{k\pm 1}{\a}, \lab{bsm1}\\[3mm]
M_{\pm}^{(k-1,k)}(\a;r,c)&=&-\xi 
{ \t_{x^8}(\xi^{-2}x^2) \ov \t_{x^8} (\xi^2 x^2) }
\nbt{c+\half+\d}{<}{k\pm1}{\a}, \lab{bsm2}
\eea
where $\d={\pi i \ov {2 \ln x}}$,  and $K'$ indicates that $r$ is
replaced by $r-1$ in the boundary weights.
The boundary S-matrix elements are thus given in terms of
the boundary Boltzmann weights (\ref{Kp0})-(\ref{Km1}) analytically continued
to shifted arguments. 
Again, this is very similar to the situation for the boundary
XXZ model \cite{JKKKM}.
The sole effect of the $\d$ is to produce the minus
sign in the shift $x^{2c} \ra -x^{2c\mp 1}$.

In order to discover discrete eigenstates
of $T_B^{(k)}(\xi)$ other than the maximal ones,
we should examine the boundary S-matrix elements in search of
simple poles in the physical strip.
We identify the boundary physical strip as $x<|\xi|<1$, and find just 
two simple boundary S-matrix poles which can
lie in this region for the allowed values of $c$.
 Firstly, we  find that there is a pole in
$M_-^{(k,k)}(\a;r,c)$ at $\xi=-x^{1-2c}$. By explicit calculation, we
have shown that this arises from a simple pole 
$\sim 1/(\xi +x^{1-2c})$ in
$\vot_-(\xi) \rbs{k}$  with residue
\be
{\rm Res}\,_{\xi=-x^{1-2c}}\, \vot_-(\xi) \ket{k,k}_B^{c}
 \sim \ket{k-1,k}_B^{c},\label{res0}. \ee
where $\ket{k-1,k}_B^{c}$ is an analytic continuation of (\ref{ansatz})
in $c$.
Equivalently, this boundary S-matrix pole arises from
a pole in  $\lbs{k} \Psi_-(\xii)$ at $\xi=-x^{1-2c}$.
 The residue at this
pole is
\be
\Res_{\xi=-x^{1-2c}}\; \lbs{k} \Psi_-(\xii) \sim \lbs{k-1}.\label{res1}\ee

The second pole that can lie in the boundary physical strip occurs in 
$M_+^{(k-1,k)}(\a;r,c)$ at $\xi=-x^{1+2c}$. This pole arise from
one in $\vot_+(\xi) \ket{k-1,k}_B^{c}$ with the following residue,
\be
{\rm Res}\,_{\xi=-x^{1+2c}}\, \vot_+(\xi) \ket{k-1,k}_B^{c}
 \sim \ket{k,k}_B^{c},\label{res2}, 
\ee
or equivalently from
\be
\Res_{\xi=-x^{1+2c}}\; \lbs{k-1} \Psi_+(\xii) \sim \lbs{k}.\label{res3}
\ee

There are no poles in $M_+^{(k,k)}(\a;r,c)$ or $M_-^{(k-1,k)}(\a;r,c)$ 
in the physical strip for either
of the parameter regions $A$ or $B$.

We shall now discuss the regions $A$ and $B$ separately and in more detail.
First consider regime $A$, in which $x^{2c}=-x^{2b}$, $-1<b<1$.
For $b>0$ there is a unique ground state with energy eigenvalue zero.
This is $|k,k\rangle^c_B$. For $0<b<\half$ the pole (\ref{res0})
in $M_-^{(k,k)}(\a;r,c)$ at $\xi=x^{1-2b}$
lies in the physical strip. The state $\rbs{k-1}$ can be interpreted as a bound
state of a `$-$ kink' with the boundary. The energy of this bound state
increases from zero at $b=0$ up to a maximum value of $\epsilon(1)$ at
$b=\half$.
This maximum value is equal to
the minimum  of the energy band of a single free kink
state (i.e. one of the form $\vot_{-}(\xi)\rbs{k}$ with
$|\xi|=1$). For $b \geq \half$
the pole moves off the physical strip, and the
state $\rbs{k-1}$ can decay by emitting a kink.
For $b<0$ the unique ground state with energy zero is $\rbs{k-1}$.
The pole in $M_+^{(k-1,k)}(\a;r,c)$ at $\xi=x^{1+2b}$ lies in the
physical strip for $-\half<b<0$. In this region $\rbs{k}$ is a bound
state of $\rbs{k-1}$ with a $+$ kink. 
Again, for $b\leq -\half$ the pole moves off the physical
strip and $\rbs{k}$ is unstable against emission of a free kink.

Now consider regime $B$,  which is parameterised by $x^{2c}=x^{2b}$,
$-1<b<1$. For $0<b<\half$ the pole (\ref{res0}) in $M_-^{(k,k)}(\a;r,c)$ at 
$\xi=-x^{1-2b}$ still lies in the physical strip, and 
the energy of the bound state $\rbs{k-1}$ is outside the
energy band of a free kink.
However, in this case, the bound state energy is
{\it greater} than the maximum 
energy $\epsilon(-1)$ of a free kink. As $b$ increases 
the energy of the bound state $\rbs{k}$ decreases and reaches
$\epsilon(-1)$ when $b$ is equal to the critical value $\half$.
For $b\geq \half$ the pole moves off the
physical strip and $\rbs{k-1}$ can decay by emission of a kink. This 
mechanism for the stability of a bound state in a lattice model
was also observed for the XXZ model in \cite{JKKKM}. 
For $b<0$,  again, and for similar reasons, there is a stable bound state
$\rbs{k}$ when $-\half<b<0$.

We remark that the normalisation  (\ref{NORM}) also
confirms the above discussion, because 
$\lbs{k}k,k\rangle_B^v$ has
has a pole at $x^{2c}=\pm x^{-1}$, and $\lbs{k-1}k-1,k\rangle_B^v$ has
a pole at $x^{2c}=\pm x$.
\setcounter{equation}{0}
\setcounter{section}{5}
\section{The Scaling Limit}
In this section we consider the $x\ra 1$ scaling limit of the bulk and
boundary S-matrices. Parameterising $\xi=x^{2\a}$, we shall identify
$\pi i \a$ with the usual rapidity parameter $\th$ in this limit. 
Our boundary physical strip becomes $0<{\rm Im}\, \th < \pi/2$ as
in \cite{Gh94}\footnote{In \cite{GZ}, the boundary physical strip was at
one point incorrectly identified as $0<{\rm Im}\, \th < \pi$. This was
corrected in \cite{Gh94}.}.
The pole at $\xi=x^{1-2|b|}$ in region A thus lies on the imaginary axis
in the $\th$ plane. It is a physical strip pole
that may be interpreted as a bound state. 
The pole at $\xi=-x^{1-2|b|}$ in region B 
moves off to infinity in the scaling limit. The energy 
of the associated state becomes infinite, as the one particle energy
band  becomes infinity high, and it should no longer be considered
as a bound state. It is for region A that we shall construct the
scaling limit of the bulk and boundary S-matrices.

\ss{A gauge transformation}
Before taking the scaling limit we carry out the following gauge
transformation:
\bea
\sg{m_4}{m_1}{m_3}{m_2}{\a} &=& - 
\left({[m_4]'[m_2]'\ov [m_1]'[m_3]'}\right)^{\a/2}
\sqrt{[m_1]'\ov [m_3]'} \;\wet{m_4}{m_1}{m_3}{m_2}{\a}, \label{gauge1} \\[5mm]
\bsg{\ep}{k-i}{\a}  
&=& \left( {[k-i+\ep]'\ov [k-i]'}\right)^{\a} M_{\ep}^{(k-i,k)}(\a), 
\label{gauge2}\\[5mm]
{\tilde{\Psi}}^{*\;(l+\ep,l)}(\xi) &=&\left( {[l+\ep]'\ov [l]'} \right)^{\a/2}
\sqrt{{[1]'\ov [l+\ep]'}}  \Psi^{*\;(l+\ep,l)}(\xi), \label{gauge3}\\[3mm]
\tilde{\Psi}_{\ep}(\xi) &=&{\tilde{\Psi}}^{*}_{-\ep}(x^2 \xi), 
\label{gauge4}\eea
where $\xi=x^{2\a}$.
The effect of this gauge transformation is to yield bulk and boundary
S-matrices $\tilde{S}$ and $\tilde{M}$ that have the conventional crossing
and boundary crossing/unitarity relations
(a rather similar gauge transformation was carried out in
reference \cite{BL90} in order
to yield a crossing symmetric S-matrix from RSOS Boltzmann weights).
Written in terms of these gauge transformed
objects, equations \mref{comm1},\mref{scond} and \mref{dscond} become,
\bea
\votg_{\varepsilon_2}(\xi_2) \votg_{\varepsilon_1}(\xi_1)
&=&\sl_{\varepsilon'_1+\varepsilon'_2=\varepsilon_1+\varepsilon_2}
\sg{L+\varepsilon_1+\varepsilon_2}{L+\varepsilon'_2}
{L+ \varepsilon_1}{L}{\a_2-\a_1}
\votg_{\varepsilon'_1}(\xi_1)
\votg_{\varepsilon'_2}(\xi_2),\label{gcomm1}\\[3mm]
\votg_{\ep}(\xi) \ket{k-i,k}_B^c &=& \Mg_{\ep}^{(k-i,k)} (\a;r,c)
\votg_{\ep}(\xii)\ket{k-i,k}_B^c \quad {\rm for}\ws i=0,1,
\label{gscond},\\[3mm]
\lbs{k-i} \tilde{\Psi}_{\ep}(\xii) &=& \lbs{k-i} \tilde{\Psi}_{\ep}(\xi) \Mg_{\ep}^{(k-i,k)}
(\a;r,c).\label{gdscond}
\eea
The S-matrix \mref{gauge1} obeys \mref{yb} and \mref{un} with the change
$W\ra \tilde{S}$. The crossing relation is modified to the more 
conventional
\be
\tilde{S}\left(\Matrix dcab{\a} \right)
=
\tilde{S}\left(\Matrix adbc{1-\a} \right)
.\\[3mm]\ee

The properties of the boundary S-matrix $\Mg$ can either be read off from the
explicit expressions \mref{bsm1} and \mref{bsm2}, or derived using the
definition \mref{gscond} and properties \mref{gcomm1} and
\mref{IIpole} of vertex operators. They are:
\bac
&&\hspace*{-10mm}\sl_{\eppp_1+\eppp_2=\ep_1+\ep_2}
\hspace*{-4mm}\sg{l+\ep_1 +\ep_2}{l+\eppp_1}{l+\ep_2}{l}{\a_1-\a_2}  
\sg{l+\ep_1+\ep_2}{l+\epp_2}{l+\eppp_1}{l}{\a_1+\a_2} 
\Mg_{\eppp_1}^{(l,k)}(\xi_1)\Mg_{\epp_2}^{(l,k)}(\xi_2) =\\[6mm] 
&&\hspace*{-10mm}\sl_{\eppp_1+\eppp_2=\ep_1+\ep_2}
\hspace*{-4mm}\sg{l+\ep_1 +\ep_2}{l+\eppp_1}{l+\ep_2}{l}{\a_1+\a_2}  
\sg{l+\ep_1+\ep_2}{l+\epp_2}{l+\eppp_1}{l}{\a_1-\a_2} 
\Mg_{\ep_2}^{(l,k)}(\xi_2) \Mg_{\eppp_1}^{(l,k)}(\xi_1),
\ea
where $\ep_1$, $\ep_2$, $\epp_1$ and $\epp_2$ are fixed with
$\ep_1+\ep_2=\epp_1+\epp_2$, and $l=k$ or $k-1$,
\bac
&&\Mg_{\ep}^{(l,k)}(\xi) \Mg_{\ep}^{(l,k)}(\xii) =1,\\[3mm]
&& \Mg_{\ep}^{(l,k)}(x^2\xii) = \sl_{\epp} \sg l {l+\epp}{l+\ep} l {\xi^2
x^{-2}} \Mg_{\epp}^{(l,k)}(\xi).
\ea

\ss{Scaling behaviour}
The S-matrices are constructed in terms of functions that
have
the following behaviour in the limit  $p \ra 1$:
\bea
\qp{p^z}{p} &\ra&(1-p)^{(1-z)} { \qp{p}{p} \ov \G (z) }, \\[2mm]
{\Theta_p(p^z) \ov \Theta_p(p^{z^{\pr}})} &\ra& {\sin(\pi z) \ov \sin(\pi
z^{\pr})}.
\eea
We shall consider the two cases $r=4$ and $r=5$, separately, and in detail. The scaling
limit of the $r=4$ model should correspond to the massive $\phi_{(1,3)}$
perturbation of the critical Ising model \cite{GZ}. That of $r=5$ should
correspond to that of the massive  $\phi_{(1,3)}$ perturbation of the
tri-critical Ising model \cite{Zam89,Ch95}.

\ss{The $r=4$ model}
For $r=4$, there are just two S-matrix elements. Reading off from
equations \mref{gcomm1}, \mref{gauge1}, and \mref{w1}-\mref{w3}, we find
\bea
\votg_+(\xi_2) \votg_-(\xi_1)
&=& \sg{2}{1}{1}{2}{\a_2-\a_1} \votg_+(\xi_1) \votg_-(\xi_2),
\quad {\rm on}\ws \cF_{2,k}\\[2mm]
\votg_-(\xi_2) \votg_+(\xi_1)
&=& \sg{1}{2}{2}{1}{\a_2-\a_1} \votg_-(\xi_1) \votg_+(\xi_2), 
\quad {\rm on}\ws \cF_{1,k},
\eea
where
\be
\sg{2}{1}{1}{2}{\a_2-\a_1}=\sg{1}{2}{2}{1}{\a_2-\a_1}=-1.\ee

This is the behaviour expected for a free
fermion theory. In order to define the scaling limit of the boundary 
S-matrix elements given by \mref{gauge2}, (\ref{bsm1}) and (\ref{bsm2}), we must 
scale $\xi=x^{2 \a}$. We consider region A in which we have the
parameterisation $x^{2c}=-x^{2b}$ with $-1<b<1$.
We find that in the limit $x\ra 1$ 
\bea
\tilde{M}_+^{(1,1)}(\a;r=4,c)=\tilde{M}_-^{(2,3)}(\a;r=4,-c)&\ra&\tan(\pi(\a/2-1/4)),
\label{fixed4},\\[3mm]
\tilde{M}_-^{(2,2)}(\a;r=4,c)=\tilde{M}_+^{(1,2)}(\a;r=4,-c)&\ra&\tan(\pi(\a/2-1/4))
\left( {\cos(\pi b) +\sin(\pi \a) \ov \cos(\pi b) -\sin(\pi \a)}\right)\label{mag4}.
\eea
We now compare these expressions with the boundary S-matrices for the
Ising model derived
in reference \cite{GZ}. We find that (\ref{fixed4}) and (\ref{mag4})
agree with the boundary S-matrices for `fixed' and `boundary magnetic
field' boundary conditions given
in (4.10) and (4.27) of \cite{GZ} if we identify the rapidity 
$\th=\pi i\a$, and the magnetic field parameter $\kappa=\cos(\pi b)$.
In addition, setting $b=0$ in (\ref{mag4}) gives
the boundary S-matrix for `free' boundary conditions as given in
(4.16) of \cite{GZ}.

\ss{The $r=5$ model}
There are just four
independent non-zero S-matrix elements in this case. 
Their scaling behaviour is given by,
\ba{lllllll}
\sg{1}{2}{2}{1}{\alpha}&=&\sg{3}{2}{2}{3}{\alpha}&\ra&
-\sqrt{2}^{1-\a} \cos({\pi \ov 4}(\a-1)) F(\a)
&=& 
-B_0(\th),\\[7mm]
\sg{3}{2}{2}{1}{\alpha}&=&\sg{1}{2}{2}{3}{\alpha}&\ra&
-\sqrt{2}^{1-\a} \cos({\pi \ov 4}(\a+1)) F(\a)
&=&-B_1(\th),\\[7mm]
\sg{2}{1}{1}{2}{\alpha}&=&\sg{2}{3}{3}{2}{\alpha}&\ra&
-\sqrt{2}^{\a}\cos({\pi \ov 4}\a) F(\a)
&=&-A_0(\th),\\[7mm]
\sg{2}{1}{3}{2}{\alpha}&=&\sg{2}{3}{1}{2}{\alpha}&\ra&
-\sqrt{2}^{\a} \sin({\pi \ov 4}\a) F(\a)
&=&-A_1(\th),
\ea
where,
\be
 F(\a) = \pl_{n >0}
{\G(n+\half)\ov \G(n-\half)}
 { 
\G(n-{\a\ov 2}) \G(n+{\a \ov 2} -\half) \ov
\G(n+{\a\ov 2}) \G(n-{\a \ov 2} +\half)}=
\left[\cos({\pi \a\ov 2})\right]^{-\half}   \exp\left(-{1\ov 4} \int_{0}^{\infty}
{dt\ov t} {\sinh(\a t) \ov \cosh^2({t / 2})}\right).
\ee
The function $F(\a)$ has the properties
\be F(\a)F(-\a)=1 \quad{\rm and}\ws
F(1-\a)=F(\a).\ee
$B_0(\th)$, $B_1(\th)$, $A_0(\th)$, $A_1(\th)$
are the S-matrix elements given in references \cite{Zam89,Ch95}.
(note however that $F(\a)$ is convergent, unlike  $S(\th)$ of references
\cite{Zam89,Ch95}). The representation of  $F(\a)$ in terms of $\G$
functions was obtained using the limiting behaviour of
$(p^z;p)_{\infty}$ discussed in Section 7.2. The integral representation
was constructed by taking the limit of the infinite sum that occurs in
$\log(p^z;p)$, and is equal to the expression given in reference 
\cite{AlZ90}. It is simple to show that the two representations agree by
making use of the integral representation of $\log(\G)$.

The symmetry of $M_{\ep}^{(k-i,k)}(\xi;r,c) = 
M_{-\ep}^{(r-1-k+i,r-k)}(\xi;r,-c)$
means that there are four independent boundary S-matrices.
They have the following scaling limits:
\ba{lllll}
\Mg_+^{(1,1)}(\a;r=5,c)&=&\Mg_-^{(3,4)}(\a;r=5,-c)&\ra&
\sqrt{2}^{\a} \tan \pi ({\a \ov 2} - {1\ov 4}) P_{\rm min}(\a),
\\[3mm]
&&&=& e^{\g \th}i\tanh ({i\pi \ov 4} -{\th \ov 2}) R_{(-1)}(\th),\\[3mm]
\Mg_-^{(3,3)}(\a;r=5,c)&=&\Mg_+^{(1,2)}(\a;r=5,-c)&\ra&
\sqrt{2}^{\a} \tan \pi ({\a \ov 2} - {1\ov 4})P_{\rm min}(\a) 
\left( {\cos(\pi b) +\sin(\pi \a) \ov \cos(\pi b) -\sin(\pi
\a)}\right),\\[3mm]
&&&=& e^{\g \th}i\tanh ({i\pi \ov 4} -{\th \ov 2}) P_{-}(\th),\\[3mm]
\Mg_{\pm}^{(2,2)}(\a;r=5,c)&=&\Mg_{\mp}^{(2,3)}(\a;r=5,-c)&\ra&\\[2mm]
&&&&\hspace*{-90mm}
\sqrt{2}^{-\a}\tan \pi ({\a \ov 2} - {1\ov 4})P_{\rm min}(\a) 
\left( {\cos(\pi b) +\sin(\pi \a) \ov \cos(\pi b) -\sin(\pi
\a)}\right)
F(\a-b-\half)F(\a+b+\half) \left(\cos(\pi({ b \ov 2}+\qu) \mp
\sin({\pi \a \ov 2})\right),\\[3mm]
&&&=& e^{\g \th} i\tanh ({i\pi \ov 4} -{\th \ov 2}) R_{\pm}(\th).
\ea
where
\be
 P_{min}(\a) = \pl_{n>0} {\G^2(n-\at) \G(n+\at+\qu) \G(n+\at-\qu) \ov
\G^2(n+\at) \G(n-\at+ \qu) \G(n-\at-\qu)}=
 \exp\left(-2 \int_{0}^{\infty}
{dt\ov t} {\sinh(\a t) \sinh^2(t/4) \ov \sinh^2(t)} \right).
\ee
$P_{min}(\a)$ has the properties 
\be P_{min}(\a)P_{min}(-\a)=1, \quad {\rm and}\ws
 P_{min}(1-\a)=\sqrt{2} \sin(\pi\a/2) F(2\a-1)P_{min}(\a).\ee
The
latter is shown most simply by using the integral representation
of both sides.
$R_{(-1)}$, $P_{-}(\th)$ and $R_{\pm}(\th)$ are the boundary
S-matrices of reference \cite{Ch95}, and $\g$ is defined in
\cite{Ch95} as $e^{2\pi i \g}=2$ (we identify $\a={\th \ov \pi i}$,
and $\pi b=\xi-{\pi \ov 2}$,
where $\xi$ is the magnetic field variable of reference
\cite{Ch95}). 

We have checked that the scaling limits of our bulk and boundary
S-matrices obey the continuum Yang-Baxter, crossing, and unitarity
conditions, and we should comment on overall factors by which our
results differ from those of references \cite{Zam89,Ch95}. 
If we wished, we could remove the overall $-$ sign in our S-matrices
by a gauge transformation. However we choose to keep it because it
introduces the $\tan \pi(\a/2 -1/4)$ factor in the boundary
S-matrices. As pointed out in reference \cite{Ch95}, we expect a
simple pole in $M_-^{(3,3)}(\a;r,c)=M_+^{(1,2)}(\a;r,-c)$ at
$\a=1/2$ when
$b=0$. The `CDD factor' $(\cos(\pi b)+\sin(\pi \a))/(\cos(\pi
b)+\sin(\pi \a))$ however has a double pole at $\a=1/2$ when $b=0$. 
So we need the factor $\tan \pi(\a/2 -1/4)$ in order
to turn the double pole into a simple pole. This situation is 
just the same as for the $r=4$ Ising model where the same factors are
present in $M_-^{(2,2)}(\a;r,c)=M+^{(1,2)}(\a;r,-c)$.
The absence of the remaining $e^{\g \th}$ in the boundary S-matrix
elements of \cite{Ch95} appears to be a simple error. Its presence is
necessary to ensure boundary crossing/unitarity.
\setcounter{equation}{0}
\setcounter{section}{6}
\section{Discussion}

In summary, we have extended the bosonization scheme of reference
\cite{LP96} to the boundary ABF models. Specifically, by making use
of bozonized vertex operators we have constructed
boundary states $\rbs{k-i}$, and the associated boundary S-matrices.
We have initially constructed $\rbs{k-i}$ in the bosonic Fock
space $\cF_{k-i,k}$, and then shown that it determines a non-zero residue class
in $\cL_{k-i,k}$ under some assumptions on the cohomological construction
proposed by Lukyanov and Pugai.
For $b>0$, $\rbs{k}$ is the vacuum
state, and $\rbs{k-1}\sim {\rm Res}\,_{\xi=-x^{1-2c}}\, \vot_-(\xi) \ket{k,k}_B^{c}$ is interpreted as 
a bound state of $\rbs{k}$ with a $-$ kink. 
This bound state is stable for $0<b<\half$.
Conversely, for  $b<0$, $\rbs{k-1}$ is the vacuum
state, and $\rbs{k}\sim{\rm Res}\,_{\xi=-x^{1+2c}}\, \vot_+(\xi) \ket{k-1,k}_B^{c}$ 
may be interpreted as a bound state of $\rbs{k-1}$ with a $+$ kink. This
bound state is stable for $-\half<b<0$.
General continous eigenstates of the boundary transfer matrix are constructed
by acting with the type II vertex operators $\vot_{\ep}$ on the
discrete states $\rbs{k-i}$.

The scaling limit of our bulk and boundary
$S$-matrices should be those associated
with the $\phi_{(1,3)}$ perturbation of the
$c=1-6/r(r-1)$ rational conformal field theories.
For $r=4$ and $r=5$ we have checked our scaling limit
against the results for these perturbed boundary conformal field
theories which are available in the literature. 

There are several directions in which this work may
be exploited and extended. 
Using our bosonized expressions
for the bounds states and vertex operators, it is possible to write
down integral expression for arbitrary equal time correlation
functions of local operators of the theory. This may be done by
a simple extension of the technique presented in \cite{JKKKM}.
Integral expressions for form factors may be derived 
in the same manner.  Difference equations for the correlation
functions and form factors may be derived following the approach
of \cite{JKKMW}. 
The results could be extended to $A_{n-1}^{(1)}$ face models using the
bosonization scheme of reference \cite{AJMP}.
\subsection*{Acknowledgments}
We wish to thank M. Jimbo, Y. Pugai and P. Dorey for interesting 
discussions. RAW would like to thank the Royal Society/JSPS/RIMS/Newton
Institute exchange scheme for providing him with funds to visit RIMS. He
would also like to thank everyone in RIMS for their hospitality during
the period in which the initial stages of this work were carried out.
\baselineskip=13pt  
\pagebreak

\baselineskip=17pt 
\appendix
\renewcommand{\thefigure}{\thesection.\arabic{figure}}
\setcounter{equation}{0}
\newpage
\section{Normal ordering relations and summation identities}
\subsection{Normal ordering relations for vertex operators}
We list the normal ordering relations for the vertex operators
discussed in Section 3. We abbreviate $AB=:AB: C$ to $AB=:\quad: C$,
and use $\z=x^{2u}$ and $z=x^{2v}$.
\bac
\vo_+(\z_1) \vo_+(\z_2) &=&:\quad:\; \z_1^{-({ r-1 \ov 2r})} 
{ \cb{x^2 \z_1/\z_2}\cb{x^{2+2r} \z_1/\z_2} \ov \cb{x^{2r}\z_1/\z_2}
 \cb{x^4 \z_1/\z_2}},\\[2mm]
\vo_+(\z)A (z)&=&:\quad:\; \z^{({r-1 \ov r})} 
{\br{z\z x^{2r-1}} \ov \br{z\z x}},\\[2mm]
A(z)\vo_+(\z)&=&:\quad:\; z^{-({r-1 \ov r})} 
{\br{(z\z)^{-1} x^{2r-1}} \ov \br{(z\z)^{-1} x}},\\[2mm]
\vo_+(\z)A (z)&=&A(z)\vo_+(\z) {[\half -u -v] \ov [\half+u+v]},\\[2mm]
A(z_1)A(z_2)&=&:\quad: z_1^{2(r-1) \ov r} (1-z_2/z_1) 
{\br{x^2 z_2/z_1} \ov \br{x^{2(r-1)}z_2/z_1} },\\[2mm]
A(z_1)A(z_2)&=&{[v_1-v_2-1] \ov [v_1-v_2+1]} A(z_2) A(z_1) ,\\[2mm]
\vot_+(\z_1) \vot_+(\z_2) &=&:\quad:\; \z_1^{-({ r \ov 2(r-1)})} 
{ \cbt{\z_1/\z_2}\cbt{x^{2(r+1)} \z_1/\z_2} \ov \cbt{x^{2r}\z_1/\z_2} \cbt{x^2 \z_1/\z_2} },\\[2mm]
\vot_+(\z)B(z)&=&:\quad:\; \z^{({r\ov r-1})} 
{\brt{z\z x^{2r-1}} \ov \brt{z\z x^{-1}}},\\[2mm]
B(z)\vot_+(\z)&=&:\quad:\; z^{-({r \ov r-1})} 
{\brt{(z\z)^{-1} x^{2r-1}} \ov \br{(z\z)^{-1} x^{-1}}},\\[2mm]
B(z)\vot_+(\z)&=& \vot_+(\z)V (z) 
{[\half -u-v]^{\pr} \ov [\half+u+v]^{\pr}} ,\\[2mm]
\vo_+(\z_1) \vot_+(\z_2) &=&:\quad:\; \z_1^{\half} 
{\qp{-x^3 \z_1/\z_2}{x^4} \ov \qp{-x\z_1/\z_2}{x^4}} ,\\[2mm]
\vot_+(\z_1) \vo_+(\z_2) &=&:\quad:\; \z_1^{\half} 
{\qp{-x^3 \z_1/\z_2}{x^4} \ov \qp{-x\z_1/\z_2}{x^4}} ,\\[2mm]
\vo_+(\z_1) \vot_+(\z_2) &=& (\z_1/\z_2)^{\half} 
{\th_{x^4}(-x\z_2/\z_1) \ov \th_{x^4}(-x\z_1/\z_2)}
\vot_+(\z_2) \vo_+(\z_1),\\[2mm]
\vo_+(\z)B(z) &=&:\quad: \zi (1+\z z),\\[2mm]
B(z)\vo_+(\z)&=&:\quad: z (1+(\z z)^{-1}),\\[2mm]
\vo_+(\z)B(z) &=&B(z)\vo_+(\z),\\[2mm]
\vot_+(\z)A(z) &=&:\quad: \zi (1+\z z),\\[2mm]
A(z)\vot_+(\z)&=&:\quad: z (1+(\z z)^{-1}),\\[2mm]
\vot_+(\z)A(z) &=&A(z)\vot_+(\z),\\[2mm]
A(z_1)B(\z_2)&=&:\quad: { 1\ov
z_1^{2}(1+xz_2/z_1)(1+x^{-1} z_2/z_1)} ,\\[2mm] 
B(z_2)A(\z_1)&=&:\quad: { 1\ov
z_2^{2}(1+xz_1/z_2)(1+x^{-1} z_1/z_21)} ,\\[2mm] 
A(z_1)B(\z_2)&=&B(z_2)A(\z_1),\\[2mm]
B(z_1)B(z_2)&=&:\quad: z_1^{2r\ov r-1} (1-z_2/z_1)
{\brt{x^{-2}z_2/z_1} \ov \brt{x^{2r} z_2/z_1}},\\[2mm]
B(z_1)B(z_2)&=&{[v_1-v_2+1]^{\pr}\ov[v_1-v_2-1]^{\pr}} 
B(z_2)B(z_1).
\ea

\subsection{Summation identities}
\bac
e^{\sl_{m>0} {z^m \ov m} {[m]_x [rm]_x \ov [2m]_x [(r-1)m]_x}}
&=& { \cbt{x^{2r} z} \cbt{x^2 z} \ov \cbt{z} \cbt{x^{2(1+r)} z}}
,\\[2mm]
e^{\sl_{m>0} {z^m \ov m} {[m]_x [(r-1)m]_x \ov [2m]_x [rm]_x }}
&=& { \cb{x^{2r} z} \cb{x^4 z} \ov \cb{x^2 z} \cb{x^{2(1+r)} z}}
,\\[2mm]
e^{\sl_{m>0} {z^m \ov m} {[m]_x \ov [2m]_x} } 
&=& {\qp{x^3 z}{x^4} \ov \qp{x z}{x^4}} ,\\[2mm]
e^{\sl_{m>0} {z^m \ov m} {[(r-1)m]_x \ov [rm]_x} } 
&=& {\br{x^{2r-1} z}\ov \br{x z}} ,\\[2mm]
e^{\sl_{m>0} {z^m \ov m} {[rm]_x \ov [(r-1)m]_x} } 
&=& {\brt{x^{2r-1} z}\ov \brt{x^{-1} z}}.
\ea
\setcounter{equation}{0}
\section{Proof of (4.6) and (4.7) for $\ep=-$ and general $k$ \label{ap2}}
In this appendix we shall give an inductive proof
of \mref{bsc1} and \mref{bsc2} for $\ep=-$ and general $k$.
The first step is to show that if we have
\be
\nbp{c}{>}{k\pm 1}{u}\Phi_{\pm}(\zeta)|k,k\rangle^c_B
=\Phi_{\pm}(\zeta^{-1})|k,k\rangle^c_B \label{given}\ee
for a given value of $k$, then it follows that
\be 
\nbpm{c}{<}{k}{u} 
\Phi_-(\zeta)|k,k+1\rangle^c_B
=\Phi_-(\zeta^{-1})|k,k+1\rangle^c_B \label{toshow}\ee
is also true for the same value of $k$.
Using (\ref{voact1}) and the vertex operator commutation relations
(\ref{exch}), we can rewrite (\ref{toshow}) as
\bac
&&
\nbpm{c}{<}{k}{u}
\left\{
\we{k}{k+1}{k+1}{k}{c+1-u}
\Phi_-(x^{2c+2})\Phi_+(\zeta)\right.\\
&&\left.+\we{k}{k-1}{k+1}{k}{c+1-u}
\Phi_+(x^{2c+2})\Phi_-(\zeta)
\right\}
|k,k\rangle^{c+1}_B=\label{4}\\[3mm]
&&\left\{
\we{k}{k+1}{k+1}{k}{c+1+u}
\Phi_-(x^{2c+2})\Phi_+(\zeta^{-1})\right.\\
&&\left.+\we{k}{k-1}{k+1}k{c+1+u}
\Phi_+(x^{2c+2})\Phi_-(\zeta^{-1})
\right\}
|k,k\rangle^{c+1}_B.\label{5}.
\ea
We can now use the identities
\bea
&&\hspace*{-7mm}
\nbpm{c}{<}{k}{u}
\we k{k+1}{k+1}k{c+1-u}
=
\nbp{c+1}{>}{k+1}{u}
\we k{k+1}{k+1}k{c+1+u}\nonumber,\\
&&\hspace*{-7mm}
\nbpm{c}{<}{k}{u}
\we k{k-1}{k+1}k{c+1-u}
=
\nbp{c+1}{>}{k-1}{u}
\we k{k-1}{k+1}k{c+1+u}\nonumber,
\eea
and the given relations (\ref{given}) to re-write 
the LHS of (\ref{4}) such that it is
identically equal to the RHS.

The next step in the proof is show
that if 
\be
\nbp{c}{<}{k-1}{u}
\Phi_-(\zeta)|k-1,k\rangle^c_B
=\Phi_-(\zeta^{-1})|k-1,k\rangle^c_B \label{toshow1}\ee
is true for a given $k$, then
\be
\nbp{c}{>}{k-1}{u}
\Phi_{-}(\zeta)|k,k\rangle^c_B
=\Phi_{-}(\zeta^{-1})|k,k\rangle^c_B \label{given1}\ee
follows.
To show this we rewrite (\ref{given1}), using (\ref{voact2}) and the 
commutation relations (\ref{comm2}), as
\bac
&&\Res_{\xi=-x^{1+2c}} \chi(\xi/\z)
\nbp c>{k-1}u
\vot_{+}(\xi)\Phi_{-}(\zeta)|k-1,k\rangle^c_B\\
&&=\Res_{\xi=-x^{1+2c}} \chi(\xi\z)
\vot_{+}(\xi) \Phi_{-}(\zeta^{-1})|k-1,k\rangle^c_B.
\ea
The equality of the two sides is then established using (\ref{given1})
and the identity
\be
\chi(\xi/\z) 
\nbp c>{k-1}u
= \chi(\xi\z) 
\nbp c<{k-1}u.
\ee

We have now completed the inductive steps. 
In Section 4 we proved (\ref{bsc1})
for $\ep=-$ and $k=1$ ((\ref{bsc1}) with $\ep=+$ and general $k$ is 
true by construction). 
The proof of \mref{bsc1} and \mref{bsc2} is complete.
\end{document}